\documentclass{article}

\usepackage{amsmath,amsfonts,amssymb}
\usepackage{fullpage}
\usepackage{graphicx} 

\newcommand{\eqdef}{\ensuremath{\stackrel{\mbox{\upshape\tiny def}}{=}}}

\newcommand{\SAT}{\mathsf{SAT}}
\newcommand{\CLIQUE}{\mathsf{CLIQUE}}

\newcommand{\eps}{\varepsilon}
\renewcommand{\epsilon}{\varepsilon}
\newcommand{\size}[1]{\lvert #1 \rvert}

\newcommand{\dist}{\mathsf{dist}}
\newcommand{\NP}{\mathsf{NP}}
\newcommand{\BPP}{\mathsf{BPP}}
\newcommand{\PC}{\mathsf{P}}
\newcommand{\B}{\Pr} 

\newsavebox{\fmbox}
\newenvironment{fmpage}[1]
     {\medskip\begin{lrbox}{\fmbox}\begin{minipage}{#1}}
     {\end{minipage}\end{lrbox}\fbox{\usebox{\fmbox}}\medskip}

\newtheorem{definition}{Definition}
\newtheorem{example}{Example}
\newtheorem{proposition}{Proposition}
\newtheorem{theorem}{Theorem}

\newtheorem{corollary}{Corollary}
\newenvironment{algo}{\leftskip=15pt \sl }{\normalsize \par }

\title{Some approximations in Model Checking and Testing }

\author{M.C. Gaudel \\ Univ Paris-Sud 
\and  R. Lassaigne \\ Univ Paris Diderot
\and  F. Magniez\\ Univ Paris Diderot
\and  M. de Rougemont\\ Univ Paris II}

\date{}

\begin{document}
\maketitle

\begin{abstract}
Model checking and testing are two areas with a similar goal: to verify that a
system satisfies a property.  They start with  different hypothesis on the systems
and develop many techniques with different notions of  approximation, when an exact
verification may be  computationally too hard. We present some  notions of
approximation with their logic and statistics backgrounds, which yield several
techniques for model checking and testing:
 {\em Bounded Model Checking, Approximate Model Checking, Approximate Black-Box
Checking, Approximate Model-based Testing and  Approximate Probabilistic Model
Checking}.
All these methods guarantee some quality and efficiency of the verification.
\end{abstract}

\newpage

\tableofcontents
\section{Introduction}
Model checking and Model-based testing are two methods for detecting 
faults in systems.
Although similar in aims, these two approaches deal with very different entities.
In model checking,  a  transition system (the {\em model}), which
describes the system, is given and checked against some required or forbidden property.
In testing, the executable system,
called the {\em Implementation Under Test } (IUT) is given as a black box:
one can only observe the behavior of the IUT on any chosen input, and then decide
whether it is acceptable or not with respect to some description of its intended
behavior.

However, in both cases the notions of models and properties play key roles: 
in model checking, the goal is to decide if a transition system
satisfies or not some given property, often given in a temporal logic,
by an automatic procedure that explores the model according to the property; 
in model-based testing, the description of the intended behavior is often given as a
transition system,
and the goal is to verify that the IUT {\em conforms} to this description.
Since the IUT is a black box, the verification process consists in using the
description model
to construct a sequence of tests, such that if the IUT passes them,
then it  conforms to the description. This is done under the assumption that
the IUT behaves as some unknown,  maybe infinite, transition system.

An intermediate activity, black box checking combines model checking and testing
as illustrated in the Figure \ref{yannakakis} below, originally set up in \cite{PVY,y04}. In this
approach, the goal 
is to verify a property of a system, given as a black box.

 We concentrate on
general results on efficient methods which guarantee some approximation, using
basic techniques from complexity theory, as some tradeoff between
feasibility and weakened objectives is needed. For example,
in model checking some abstractions are made
on the transition system according to the property to be checked. 
In testing, some assumptions are made on the IUT,  like an upper bound on the number
of states,
or the uniformity of behavior on some input domain. These assumptions express
the gap
between the success of a finite test campaign and conformance. 
These abstractions or assumptions are specific to a given situation and
 generally do not fully guarantee the correctness.

\begin{figure}[h]
\centerline{\includegraphics[height=4.2cm]{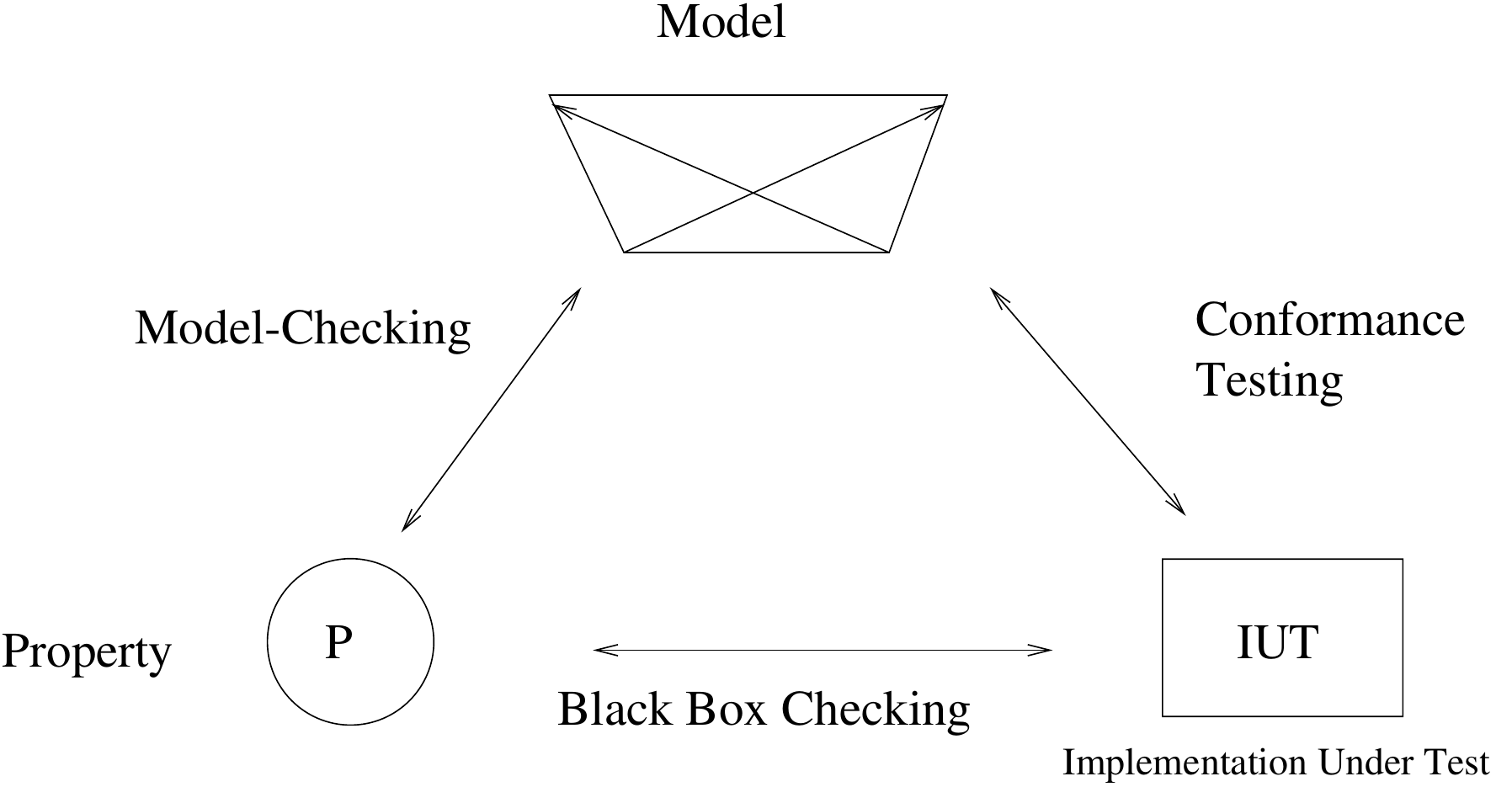}}
\caption{Model checking,   black box checking and testing.\label{yannakakis}}
\end{figure}

This paper presents different notions of approximation which may be used
 in the context of  model checking and testing. Current methods such as bounded
model checking and abstraction, and most testing methods  use some notions of
approximation but it is  difficult to quantify their quality. In this framework,
hard problems for some complexity measure may become easier when both randomization
and approximation are used. Randomization alone, i.e. algorithms of 
the class $\BPP$ may not suffice to obtain efficient solutions, as $\BPP$ may be
equal to $\PC$. 
Approximate randomized algorithms  trade 
approximation with
efficiency, i.e. relax the correctness property in order to develop efficient
methods which guarantee the quality of the approximation. This paper emphasizes the
variety of possible approximations which may lead to efficient verification methods,
in time polynomial or  logarithmic in the size of the domain, or
constant (independent of the size of the domain), and the connections between some of them.

Section~2 sets the framework for model checking and model-based testing. 
Section~3 introduces two kinds of approximations: approximate techniques for  satisfiability, equivalence and counting problems, and
randomized techniques for the approximate versions of satisfiability and equivalence problems. {\em Abstraction} as a method to approximate a model checking problem, {\em Uniform generation and Counting}, and  {\em Learning} are introduced in section 3.1.
  Property testing, the basic approach to approximate decision and equivalence problems, as well as
statistical learning
are defined in Section~3.2.
Section~4 describes the five different types of approximation that we review in this
paper, based on the logic and statistics tools of Section~3 for model checking and
testing:

\begin{enumerate}
\item {\em Bounded Model Checking} where the computation paths are bounded (Section 4.1),
\item  {\em Approximate Model Checking} where we use two distinct approximations: the proportion of inputs which separate the model and the property, and 
some edit distance between a model and a property  (Section 4.2),
\item {\em  Approximate Black Box Checking} where one approximately learns a model (Section
4.3),
\item  {\em Approximate Model-based Testing} where one finds tests which approximately
satisfy some coverage criterium (Section 4.4),
\item  {\em Approximate Probabilistic Model Checking} where one approximates the
probabilities of satisfying formulas (Section 4.5).
\end{enumerate}

The methods we describe guarantee some quality of  approximation and a complexity
which ranges from polynomial in the size of the model, polynomial in the size of
the representation of the model, to constant time:
\begin{enumerate}
\item In bounded model checking, some upper bounds on the execution paths to witness
some error are stated for some class of formulas. The method is polynomial in the
size of the model.
 \item In approximate model checking, the methods  guarantee with high probability that we discover some errors.
We use two criteria. In the first approach, if the  density of errors is larger than $\eps$, Monte Carlo methods find them with high probabilities in polynomial time.  In the second approach, if the distance of  the inputs to the property is larger  than $\eps$, an error will be found 
with high probability. The time complexity is constant, i.e. independent of the size of
the model but dependent on $\eps$.

\item In approximate black box checking, learning techniques construct a model which can be compared with a property.
Some intermediate steps, such as model checking are exponential in the  size of the model. These steps can be approximated
using the previous approximate model checking and
  guarantee that the  model is
$\eps$-close to the IUT after $N$ samples, using learning techniques which depend on $\eps$.

\item In approximate model-based testing, a coverage criterium is satisfied with
high probability which depends on the number of tests. The method is polynomial in
the size of the representation.

\item In approximate probabilistic model checking,  the estimated probabilities of
satisfying formulas are close to the real ones. The method is polynomial in the size
of a succinct representation.
\end{enumerate}

The paper focuses on approximate and randomized algorithms in model checking  and
model-based testing. Some common techniques and methods are pointed out.
Not surprisingly the use of model checking techniques for model-based test 
generation has been extensively studied. Although of primary interest, this 
subject is not treated in this
 paper.

 We believe that this survey will encourage some cross-fertilization
and new tools both for  approximate
and probabilistic model checking, and for randomized model-based testing. 
 
\section{Classical methods in model checking and testing}

Let $P$ be a finite set of atomic propositions, and ${\cal P}(P)$ the power set
of $P$.
A {\em Transition System}, or a Kripke structure, is a structure
 $\mathcal{M}=(S,s_0,R,L)$ where
 $S$ is a finite set of states, $s_0 \in S$ is the  initial state,
 $R \subseteq S\times S $ is the transition relation between states  
and $L: ~~ S \rightarrow {\cal P}(P)$ is the labelling function. This function
assigns labels to states such that if $p \in P$ is an atomic proposition, then
 $\mathcal{M},s \models p$, i.e. $s$ satisfies $p$ if $p \in L(s)$. Unless otherwise
stated, the {\em size of $\mathcal{M}$} is $|S|$, the size of $S$.

A {\em Labelled Transition System} on a finite alphabet
  $I$ is a  structure 
$\mathcal{L}=(S,s_0,I,R,L)$ where $S,s_0,L$  are as before and 
$R \subseteq S\times I \times S$. The transitions  have labels in $I$.
A run on a word $w \in I^*$ is a sequence of states $s_0,s_1,....,s_n$ such that
$(s_i,w_i,s_{i+1}) \in R$ for $i=0,...,n-1$.

A {\em Finite State Machine} (FSM) is a  structure
$\mathcal{T}=(S,s_0,I,O,R)$ with input alphabet $I$ and output 
alphabet $O$ and $R \subseteq S\times I \times O \times S$. An output word  $t \in O^*$
is produced by an input word $w \in I^*$ of the FSM
  if there is a run, also called a trace,  on $w$, i.e.
 a sequence of states $s_0,s_1,...,s_n$ such that
$(s_i,w_i,t_i,s_{i+1}) \in R$ for $i=0,...,n-1$. The input/output relation is the pair
$(w,t)$ when $t$ is produced by $w$. An FSM is {\em deterministic } if there is a
function $\delta$ such that $\delta(s_i,w_i)=(t_i,s_{i+1})~ {\rm iff}~
(s_i,w_i,t_i,s_{i+1}) \in R$. There may be a label function $L$ on the states, in
some cases.

Other important models are introduced later. An {\em Extended Finite State Machine} (EFSM),
introduced 
in section \ref{efsm}, assigns variables and
their values to states and is a succinct representation of a much larger FSM.
Transitions assume guards and define updates on the variables.
A   {\em B\"{u}chi automaton}, introduced in section \ref{ba}, generalizes classical
automata, i.e. FSM with no output but with accepting states, to infinite words.
In order to consider probabilistic systems, we introduce  {\em Probabilistic Transition
Systems}  and {\em Concurrent  Probabilistic Systems} in section \ref{pts}.\\

\subsection{Model checking}\label{mc}

Consider a transition system $\mathcal{M}=(S,s_0,R,L)$ and a temporal property expressed by a
formula $\psi$ of {\em Linear Temporal Logic} (LTL) or {\em Computation Tree Logic} (CTL and CTL$^*$). 
The {\em Model Checking} problem is to decide whether
$\mathcal{M} \models \psi$, i.e. if the system $\mathcal{M}$ satisfies the property defined
by $\psi$, and to give a counterexample if the answer is negative.

 In linear temporal logic LTL, formulas are composed from the set of atomic propositions
 using the boolean connectives and the main temporal operators  $\mathbf{X}$ 
(\textit{next time}) and  $\mathbf{U}$ (\textit{until}).
In order to analyze the sequential behavior of a transition system $\mathcal{M}$,  
LTL formulas are interpreted over runs or execution paths of the transition system ${\cal M}$. 
A path $\sigma $ is an infinite sequence of states $(s_0,s_1,\dots,s_i,\dots )$ such that 
$(s_i,s_{i+1}) \in R$ for all $i\geq 0$. We note $\sigma^i$ the path $(s_i,s_{i+1},\dots)$.
The interpretation of LTL formulas  are defined by:

\begin{itemize}
\item if $p \in P$ then $\mathcal{M}, \sigma \models p$ iff $p \in L(s_0)$,
\item $\mathcal{M}, \sigma \models \neg \psi$ iff $\mathcal{M}, \sigma \not\models \psi$,
\item $\mathcal{M}, \sigma \models \varphi \wedge \psi$ iff $\mathcal{M}, \sigma \models \varphi $ and $\mathcal{M}, \sigma \models \psi$,
\item $\mathcal{M}, \sigma \models \varphi \vee \psi$ iff $\mathcal{M}, \sigma \models \varphi $ or $\mathcal{M}, \sigma \models \psi$,
\item $\mathcal{M}, \sigma \models\mathbf{X} \psi$ iff $\mathcal{M}, \sigma^1 \models \psi$,
\item $\mathcal{M}, \sigma \models \varphi \mathbf{U} \psi$ iff there exists $i\geq 0$
such that $\mathcal{M}, \sigma^i \models \psi$ and for each $0\leq j<i$, 
$\mathcal{M}, \sigma^j \models \varphi$,
\end{itemize}

The usual auxiliary operators $\mathbf{F}$ (\textit{eventually}) and $\mathbf{G}$ (\textit{globally}) can also be defined: $true \equiv p \vee \neg p$ for some arbitrary $p \in P$, $\mathbf{F}\psi \equiv true \mathbf{U} \psi$ and $\mathbf{G} \psi \equiv \neg \mathbf{F} \neg \psi$.

In Computation Tree Logic CTL$^*$, general formulas combine states and paths formulas.
\begin{enumerate}
\item A state formula is either
 \begin{itemize}
 \item $p$ if $p$ is an atomic proposition, or
 \item $\neg F$, $F \wedge G$ or $F \vee G$ where $F$ and $G$ are state formulas, or
 \item $\exists \varphi$ or $\forall \varphi$ where $\varphi$ is a path formula.
 \end{itemize}
\item A path formula is either
  \begin{itemize}
  \item a state formula, or
  \item $\neg \varphi$, $\varphi \wedge \psi$, $\varphi \vee \psi$, $\mathbf{X} \varphi$ or $\varphi \mathbf{U} \psi$ where $\varphi$ and $\psi$ are path formulas.
   \end{itemize}
 \end{enumerate}
    
State formulas are interpreted on states of the transition system. The meaning of path quantifiers is defined by:
given $\mathcal{M}$ and $s\in S$, we say that $\mathcal{M},s \models \exists \psi$
(resp. $\mathcal{M},s \models \forall \psi$)  if there exists a path $\pi$ starting in $s$ which satisfies $\psi$ (resp. all paths $\pi$ starting in $s$  satisfy $\psi)$. 

In CTL, each of the temporal operators $\mathbf{X}$ and $\mathbf{U}$ must be immediately preceded by a path quantifier. LTL can be also considered as the fragment of CTL$^*$ formulas of the form $\forall \varphi$ where $\varphi$ is a path formula in which the only state subformulas are atomic propositions . It can be shown that the three temporal logics  CTL$^*$, CTL and LTL have different expressive powers.

 
The first model checking algorithms enumerated the reachable states of the system in order to check the correctness of a given specification expressed by an LTL or CTL formula. The time complexity of these algorithms was linear in the size of the model and of the formula for CTL, and linear in the size of the model and exponential in the size of the formula for LTL. The specification can usually 
be expressed by a formula of small size, so the complexity depends in a crucial way on the model's size. Unfortunately, the representation of a protocol or of a program with boolean variables by a transition system   illustrates the {\em state explosion phenomenon}: the number of states of the model is exponential in the  number of variables. During the last twenty years, different techniques have been used to reduce the complexity of temporal logic model checking:

\begin{itemize}
\item automata theory and on-the-fly model construction,
\item symbolic model checking and representation by ordered binary decision diagram (OBDD),
\item symbolic model checking using propositional satisfiability (SAT) solvers.
\end{itemize}

\subsubsection{Automata approach}\label{ba}
This approach to verification is based on an intimate connection between linear temporal logic and automata theory for infinite words which was first explicitly discussed in \cite{WVS83}.
The basic idea is to associate with each linear temporal logic formula a finite  
automaton over infinite words that accepts exactly all the runs that satisfy the 
formula. This enables the reduction of decision problems such as satifiability and model 
checking to known automata-theoretic problems.

 A \textit{nondeterministic B\"{u}chi} automaton is a tuple $\mathcal{A}= (\Sigma ,S, S_0, \delta ,F)$, 
where
\begin{itemize}  
\item $\Sigma$ is a finite alphabet,
\item $S$ is a finite set of states,
\item $S_0\subseteq S$ is a set of initial states,
\item $\delta : S\times \Sigma \longrightarrow 2^S$ is a transition function, and
\item $F\subseteq S$ is a set of final states.
\end{itemize}

The automaton $\mathcal{A}$ is deterministic if  $|\delta (s,a)| = 1$ for all states $s\in S$, for all $a\in \Sigma$, and if $|S_0| = 1$.

A run of $\mathcal{A}$ over a infinite word $w=a_0a_1\dots a_i\dots$ is a sequence $r=s_0s_1\dots s_i\dots$  where $s_0\in S_0$ and $s_{i+1} \in\delta (s_i,a_i)$ for all $i\geq 0$. The limit of a run $r=s_0s_1\dots s_i\dots$ is the set  $lim(r) = \{s | s=s_i \mbox{ for infinitely many} ~i\}$. 
A run $r$ is \textit{accepting} if $lim(r) \cap F \neq \emptyset$. 
An infinite word $w$ is accepted by $\mathcal{A}$ if there is an accepting run of $\mathcal{A}$ over $w$. The language of $\mathcal{A}$, denoted by the regular language $L(\mathcal{A})$, is the set of infinite words accepted by $\mathcal{A}$. 
For any LTL formula $\varphi$, there exists a nondeterministic B\"{u}chi automaton $\mathcal{A}_{\varphi}$ such that the set of words satisfying $\varphi$ is the regular language $L(\mathcal{A})_{\varphi}$ and that can be constructed in time and space $O(|\varphi |.2^{|\varphi |})$.
Moreover any transition system  $\mathcal{M}$ can be viewed as a B\"{u}chi automaton
$\mathcal{A}_{\mathcal{M}}$. Thus model checking can be reduced to the comparison of two infinite regular languages and to the emptiness problem for regular languages \cite{vw} :
$\mathcal{M} \models \varphi$ iff $L(\mathcal{A}_{\mathcal{M}}) \subseteq L(\mathcal{A}_{\varphi})$ iff
$L(\mathcal{A}_{\mathcal{M}}) \cap  L(\mathcal{A}_{\neg\varphi}) = \emptyset$ iff  
$L(\mathcal{A}_{\mathcal{M}} \times \mathcal{A}_{\neg\varphi}) = \emptyset$.

In \cite{vw}, the authors prove that LTL model checking can be decided in time $O(|\mathcal{M}|.2^{|\varphi|})$ and in space $O((log |\mathcal{M}|+|\varphi |)^2)$, that is a refinement of the result in \cite{sc}, which says that LTL model checking is PSPACE-complete. One can remark that a time upper bound that is linear in the size of the model and exponential in the size of the 
formula is considered as reasonable, since the specification is usually rather short. However, the main  problem is the state explosion phenomenon due to the representation of a protocol or of a program to check, by a transition system.

The automata approach can be useful in practice for instance when the transition system is given as a product of small components $\mathcal{M}_1, \dots, \mathcal{M}_k$. The model checking can be done without building the product automaton, using space $O((log |\mathcal{M}_1| + \dots + log |\mathcal{M}_k|)^2)$ which is usually much less than the space needed to store the product automaton. In \cite{gpvw}, the authors describe a tableau-based algorithm for obtaining an automaton from an LTL formula. Technically, the algorithm translates an LTL formula into a generalized B\"{u}chi automaton using a depth-first search. A simple transformation of this automaton yields a classical B\"{u}chi automaton for which the emptiness check can be done using a cycle detection scheme. The result is a verification algorithm in which both the transition model and the property automaton are constructed on-the-fly during a depth-first search that checks for emptiness. This algorithm is adopted in 
 the model checker SPIN \cite{spin}.

\subsubsection{OBDD approach}

In symbolic model checking  \cite{BCMD92,macm}, the transition relation is coded symbolically as a boolean expression, rather than expicitly as the edges of a graph. A major breakthrough was achieved by the introduction of OBDD's as a data structure for representing boolean expressions in the model checking procedure.

An  ordered binary decision diagram (OBDD) is a data structure
which can encode an arbitrary relation or boolean function on a finite domain.
Given a linear  order $<$ on the variables, it is a binary decision diagram, i.e.a
 directed acyclic graph with exactly one root, two sinks, labelled by
the constants $1$ and $0$, such that
each non-sink node is labelled by a variable $x_i$, and has two outgoing
 edges which are labelled by $1$ ($1$-edge) and $0$ ($0$-edge), respectively. 
The order, in which the variables appear on a path in the graph, is
 consistent with the variable order $<$, i.e. for each edge connecting a
 node labelled by $x_i$ to a node labelled by $x_j$, we have  $x_i < x_j$.\\

\begin{figure}[h]
\begin{center}

\centerline{\includegraphics[width=8cm]{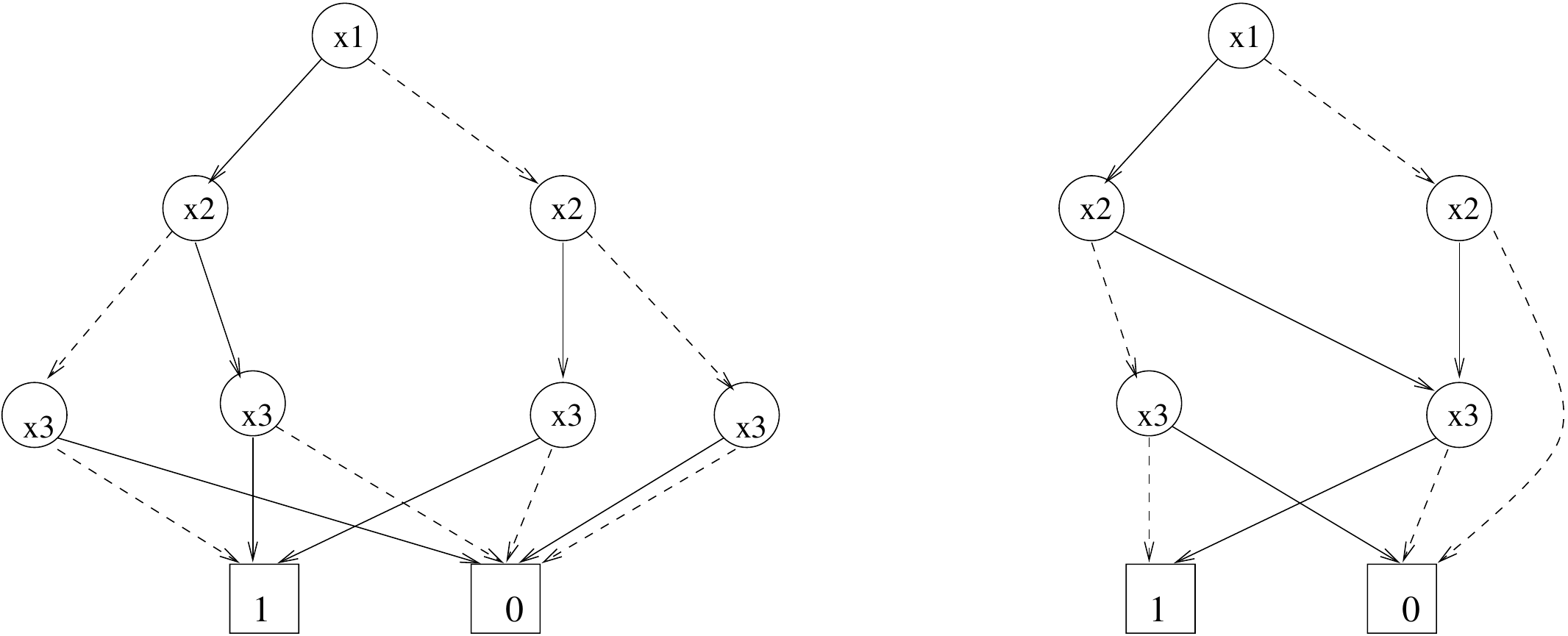}} 
\caption{Two OBDDs for a function $f: \{0,1\}^3 \rightarrow \{0,1\}$.}
\label{obdd}
\end{center}
\end{figure}

\begin{figure}[h]\label{union}
\begin{center}

\centerline{\includegraphics[width=10cm]{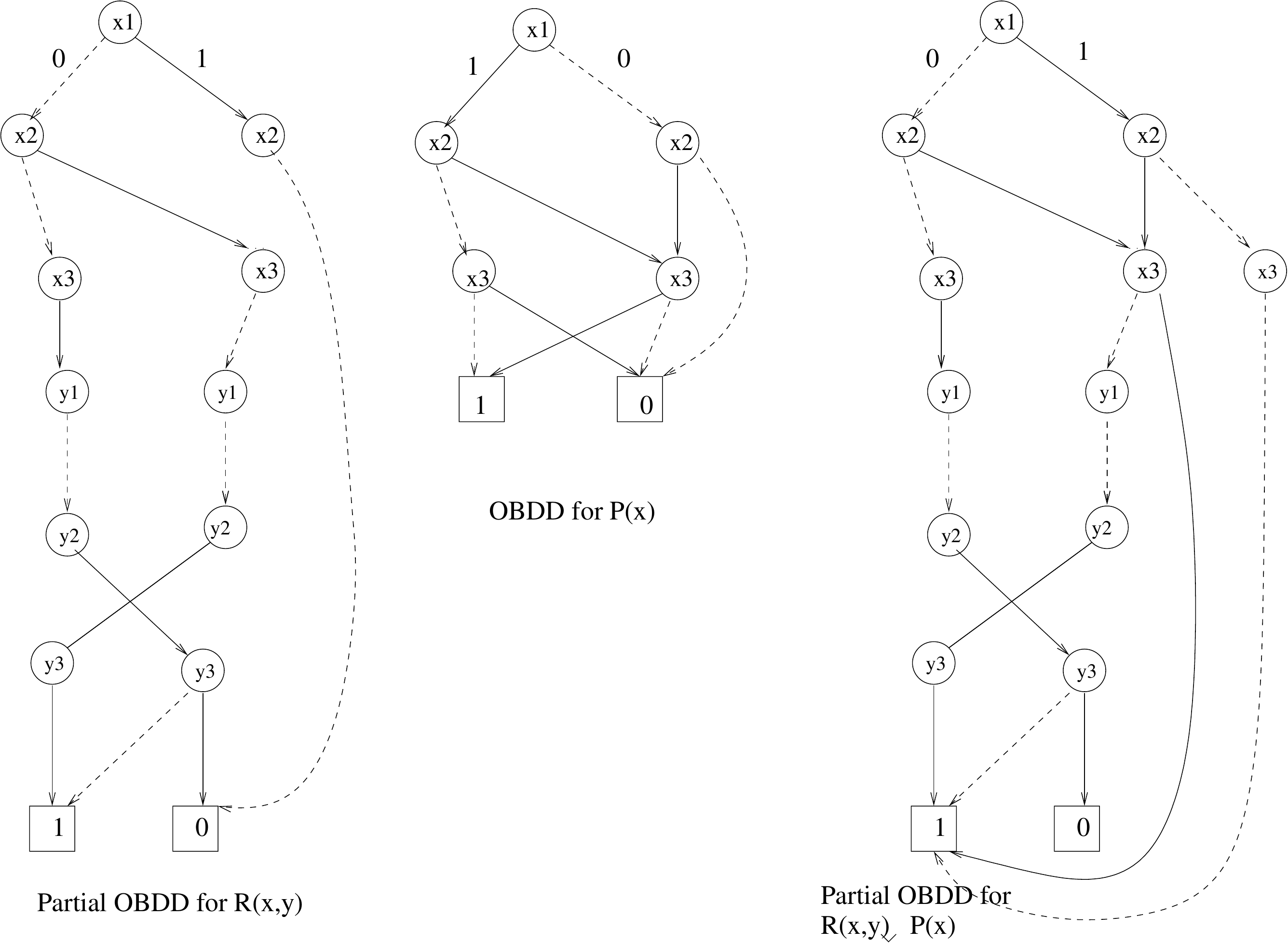}} 
\caption{The construction of an OBDD for $R(x,y) \vee P(x)$.}
\label{obdd}
\end{center}
\end{figure}

Let us start with an  OBDD representation of the relations $R$  of
$\mathcal{M}$, the
transition relation, and of each unary relation $P(x)$  describing states which satisfy the
atomic propositions $p$. Given a CTL formula, one constructs by induction on its
syntactic structure, an OBDD for the unary relation defining
the states where it is true, and we can then decide if $\mathcal{M} \models \psi$.
Figure \ref{union} describes the construction of an OBDD for $R(x,y) \vee P(x)$ from an OBDD for $R(x,y)$ and an OBDD for $P(x)$.
Each variable $x$ is decomposed in a sequence of boolean variables. In our example $x_1,x_2,x_3$ represent $x$ and similarly for $y$. The order of the variables is $x_1,x_2,x_3,y_1,y_2,y_3$ in our example. Figure \ref{union} presents a partial decision tree: the dotted line corresponds to $x_i=0$ and the standard line corresponds to $x_i=1$. The tree is partial to make it readable, and missing edges lead to $0$. 
The main drawback is that the OBDD can be exponentially large, even for simple
formulas \cite{b91}. The good choice of the order  on the variables is important, as the size of the OBDD may vary exponentially if we change the order.


\subsubsection{SAT approach}

Symbolic model checking and symbolic reachability analysis can be reduced to the satisfiability problem for propositional formulas 
\cite{BCCZ99,ABE00}. These reductions will be explained in the section 4.1: bounded and unbounded model checking. In the following, we recall the quest for efficient satisfiability solvers which has been the subject of an intensive research during the last twenty years.

Given a propositional formula which is presented in a Conjunctive Normal
 Form (CNF), the goal is to find a positive assignment of the formula.
 Recall that,  a CNF is a conjunction of one or more clauses
  $C_1\wedge C_2\wedge C_3\wedge\ldots$, where each
  clause is a disjunction of one or more literals, $C_1=x_1\vee \bar x_2\vee \bar x_5
  \vee x_7$,
  $C_2=\bar x_3\vee x_7$, $C_3=\ldots$.
  A literal is either the positive or the negative occurrence of a propositional
  variable,
  for instance $x_2$ and $\bar x_2$ are the two literals for the variable $x_2$.

  Due to the NP-completeness of SAT, it is unlikely that
  there exists any polynomial time solution. However,
  NP-completeness does not exclude the possibility of finding
   algorithms that are efficient enough for solving many interesting SAT instances.
    This was the motivation for the development of several successful algorithms \cite{DBLP:conf/cav/ZhangM02}.

    An original important algorithm for solving SAT, due to \cite{dp}, 
    is based on two simplification rules and one resolution rule. As this algorithm suffers from a
    memory explosion,  \cite{dll} proposed a modified version (DPLL) which performs a branching search with backtracking, in order to reduce the memory space required by the solver.

    \cite{grasp} proposed an iterative version of DPLL,
    that is a branch and search algorithm. Most of the modern SAT solvers
    are designed in this manner and the main components of these
    algorithms are: 
    \begin{itemize}
    \item a decision process to extend the current assignment  to an unassigned
    variable; this decision is usually based on \textit{branching} heuristics,
    \item a deduction process to propagate the logical consequences of an assignment 
    to all clauses of the SAT formula; this step is called \textit{Boolean Constraint 
    Propagation} (BCP),
    \item a \textit{conflict analysis} which may lead to the identification of one or 
    more unsatisfied clauses, called conflicting clauses,
    \item a \textit{backtracking} process to undo the current assignment and to try 
    another one.
    \end{itemize}

    In a SAT solver, the BCP step is to propagate the consequences of the current 
    variable assignment to the clauses. In CHAFF \cite{chaff}, Moskewicz et al. proposed a BCP
    algorithm called two-literal watching with lazy update. Since the breakthrough of CHAFF, most effort  in the design of efficient SAT solvers has been focused  on efficient BCP, the heart of all modern SAT solvers.

    An additional technique named  \textit{Random restart}
    was proposed to cope with the following phenomenon: two instances with the same clauses
     but different variable orders may  require different times by a  SAT solver. Experiments show that a random restart can increase the robustness of SAT solvers and this technique is
     applied in modern SAT solvers such as RSTART \cite{pd}, TiniSAT \cite{hj07} and PicoSAT \cite{ab}. This technique, for example the nested restart scheme used by PicoSAT,  is inspired by the work of M. Luby et al. \cite{lsz}.

Another significant extension of DPLL is clause learning: when there is a
conflict after some propagation, and there are still some branches to be
searched, the cause of the conflict is analysed and added as a new clause
before backtracking and continuing the search \cite{BKS03}. Various
learning schemes have been proposed \cite{AS09} to derive the new clauses.
Combined with non chronological backtracking and random restart these
techniques are currently the basis of modern SAT-solvers, and the origin
of the spectacular increase of their performance.

\subsection{Verification of  probabilistic systems}\label{pts}

In this section, we consider systems modeled either as finite discrete time
Markov chains or as Markov models enriched with a nondeterministic
behavior. In the following, the former systems will be denoted by
probabilistic sytems and the latter by concurrent probabilistic
sytems. 
A Discrete Time Markov Chain (DTMC) is a pair $(S,M)$ where
$S$ is a finite or countable set of states and $M : S \times S
\rightarrow [0,1]$ is the stochastic matrix giving the transition probabilities,
i.e. for all
$s \in S$, $\sum_{t \in S} M (s,t) =1$. In the following, the set of
states $S$ is finite.

\begin{definition}
  A {\em probabilistic transition system} ($PTS$) is a structure $\mathcal{M}_p =
  (S,s_0,M,L)$ given by a Discrete Time Markov chain $(S,M)$ with an
  initial state
  $s_0$ and a function $L:S\rightarrow \mathcal{P}(P)$ labeling each
  state with a set of atomic propositions in $P$.
\end{definition}
 
A \textit{path} $\sigma $ is a finite or infinite sequence of
 states $(s_0,s_1,\dots,s_i,\dots )$ such that $P(s_i,s_{i+1})$ $>0$
 for all $i\geq 0$. We denote by $Path(s)$ the set of paths whose first state
 is $s$. For each structure ${\cal M}$ and state $s$, it is possible to define
 a probability measure $Prob$ on the set $Path(s)$. For any finite
 path $\pi =(s_0,s_1,\dots, s_n)$, the measure is defined by:
 $$Prob(\{\sigma ~:~ \sigma \mbox{ is a path with prefix } ~\pi\}
 ) = \prod_{i=1}^n
 M(s_{i-1},s_i)$$
 This measure can be extended uniquely to the Borel
 family of sets generated by the sets $\{\sigma ~:~  \pi \mbox{ is a
 prefix of } \sigma\}$ where $\pi$ is a finite path.
 In \cite{va}, it is shown that for any $LTL$ formula $\psi$,
 probabilistic transition system ${\cal M}$ and state $s$, the set of
 paths $\{\sigma ~:~ {\sigma_0 = s \mbox{ and } \cal M},\sigma \models
 \psi\}$ is measurable. We denote by $Prob[\psi]$ the measure of this
 set and by $Prob_k [\psi]$ the probability
measure associated to the probabilistic space of execution paths of finite
length $k$. 

\subsubsection{Qualitative verification}

 We say that a probabilistic transition sytem ${\cal M}_p$
 satisfies the formula $\psi$ if $Prob[\psi] = 1$,  i.e. if almost all
 paths in ${\cal M}$, whose origin is the initial state, satisfy
 $\psi$. The first application of verification methods to probabilistic systems
consisted in checking if temporal properties are satisfied with
probability $1$ by a finite discrete time Markov chain or by a concurrent probabilistic
sytem.  \cite{va} presented the first method to verify if
a linear time temporal property is satisfied by almost all
computations of a concurrent probabilistic system. However, this
automata-theoretic method is  doubly exponential in
the size of the formula.

The complexity  was later
addressed in \cite{cy}. A new model
checking method for probabilistic systems was introduced, whose
complexity was  polynomial in the size of the system
and exponential in the size of the formula. For concurrent
probabilistic systems they presented an automata-theoretic approach
which improved on Vardi's method by a single exponential in the
size of the formula.

\subsubsection{Quantitative verification}

The  \cite{cy} method  allows to compute
the probability that a probabilistic system satisfies some given
linear time temporal formula. 

\begin{theorem}(\cite{cy})
 The satisfaction of a $LTL$ formula $\phi$ by a probabilistic
  transition sytem ${\cal M}_p$ can be decided in time linear in the size of
  ${\cal M}_p$ and exponential in the size of $\phi$, and in space
  polylogarithmic in the size of ${\cal M}_p$ and polynomial in the size of
  $\phi$. The probability $Prob[\phi]$ can be computed in time
  polynomial in size of ${\cal M}_p$ and exponential in size of $\phi$.
\end{theorem}

 A temporal logic for the specification
of quantitative properties, which refer to a bound of the probability
of satisfaction of a formula, was given in
\cite{hj}. The authors introduced the logic PCTL, which is an
extension of branching time temporal logic CTL with some probabilistic
quantifiers. A model checking algorithm was also presented: the
computation of probabilities for formulas involving probabilistic
quantification is performed by solving a linear system of equations,
the size of which is the model size.\\
A model checking method for concurrent probabilistic systems against
PCTL and PCTL$^*$ (the standard extension of PCTL) properties is
given in  \cite{bda}.  Probabilities are computed
by solving an optimisation problem over system of linear inequalities,
rather than linear equations as in \cite{hj}. The algorithm for the
verification of PCTL$^*$ is obtained by a reduction to the PCTL
model checking problem using a transformation of both the formula and
the probabilistic concurrent system. Model checking of PCTL formulas
is shown to be polynomial in the size of the system and linear in the
size of the formula, while PCTL$^*$ verification is polynomial in the
size of the system and doubly exponential in the size of the formula.

In order to illustrate space complexity problems, we mention the main
model checking tool for the verification of quantitative properties. 
 The probabilistic model checker PRISM \cite{dakn,hknp} was designed by the
 Kwiatkowska's team  and allows to check
PCTL formulas on probabilistic or concurrent probabilistic systems.
 This tool uses
extensions of OBDDs called Multi-Terminal Binary Decision Diagrams
(MTBDDs) to represent Markov transition matrices, and 
 classical techniques for the resolution of linear systems. 
 Numerous classical protocols represented as probabilistic or
concurrent probabilistic systems have been successfully verified by
PRISM. But experimental results are often limited by the exponential
blow up of space needed to represent the transition matrices and to
solve linear systems of equations or inequations. In this context,
it is natural to ask the question: {\em can probabilistic verification
be efficiently approximated?} We study in Section \ref{APMC} some possible
answers for  probabilistic transition systems and linear time temporal
logic.

\subsection{Model-based testing}
\label{MBT}

Given some executable implementation under test and some description of its
expected behavior, the IUT is submitted to experiments based on the description. 
The goal is to (partially) check that the IUT is conforming to the description. 
As we explore links and similarities with model checking, we focus on descriptions 
defined in terms of finite and infinite state machines, transitions systems, and
automata.  The corresponding testing methods are  called {\em Model-based
Testing}.

Model-based testing has received a lot of attention  and is
now a well established discipline (see for instance \cite{LY,BT,B05}).
Most approaches have focused on the deterministic derivation from a
finite model of some so-called checking sequence, or of some
complete set of test sequences, that ensure conformance of
the IUT with respect to the model.
However, in very large models, such approaches are not practicable and
some selection strategy must be applied to obtain test sets of reasonable
size.  A popular selection criterion is the {\em transition coverage}.  Other
selection methods rely on the statement of some test purpose or on random choices
among input sequences or traces.
 
 \subsubsection{Testing based on finite state machines \label{test}}

As in \cite{LY}, we first consider testing methods based on deterministic FSMs: 
instead of $\mathcal{T}=(S,s_0,I,O,R)$ where 
$R \subseteq  S \times I \times O \times S$, 
we have 
$\mathcal{F} = (S, I, O, \delta, \lambda)$. 
where $\delta$ and $\lambda$ are functions from $S \times I$ into $S$, and from $S
\times I $ into $O$, respectively.
There is not always an initial state.
Functions  $\delta$ and $\lambda$ can be extended in a canonic way to sequences of
inputs:  
$\delta^*$ is from $S \times I^*$ into $S^*$and $\lambda^*$ is from $S \times I^* $
into $O^*$.

The testing problem addressed in this subsection is: given a deterministic
specification FSM $A$, and an IUT that is supposed to behave as some unknown
deterministic FSM $B$, how to test that $B$ is equivalent to $A$ via inputs
submitted to the IUT and outputs observed from the IUT?
The specification FSM must be strongly connected, i.e., there is a path between
every pair of states: this is necessary for designing test experiments that reach
every specified state.

Equivalence of FSMs is defined as follows.
Two states $s_i$  and $s_j$  are equivalent if and only if for every input sequence,
the FSMs will produce the same output sequence, i.e., for every input sequence
$\sigma$, $\lambda^*(s_i , \sigma) = \lambda^*(s_j , \sigma)$.
$\mathcal{F}$ and $\mathcal{F}'$ are equivalent if and only for every state in
$\mathcal{F}$ there is a corresponding equivalent state in $\mathcal{F}'$, and vice
versa. When $\mathcal{F}$ and $\mathcal{F}'$ have the same number of states, this
notion is the same as isomorphism.
Given an FSM, there are well-known polynomial algorithms for constructing a
minimized (reduced) FSM  equivalent to the given FSM, where there are no equivalent
states. 
The reduced FSM is unique up to isomorphism. The specification FSM is supposed to be
reduced before any testing method is used. 

Any test method is based on some assumption on the IUT called \emph{testability
hypotheses}. 
An example of a non testable IUT would be a ``demonic'' one that would behave well
during some test experiments and change its behavior afterwards.
Examples of classical testability hypotheses, when the test is based on finite state
machine descriptions, are:
\begin{itemize}

\item The IUT behaves as some (unknown) finite state machine. 

\item The implementation machine does not change during the experiments.  

\item It has the same input alphabet as the specification FSM. 

\item It has a known number of states greater or equal to the specification FSM. 

\end{itemize}

This last and strong hypothesis is necessary to develop testing methods that reach a
conclusion after a finite number of experiments. 
In the sequel, as most authors, we develop the case where the IUT has the same
number of states as the specification FSM.
Then we give some hints on the case where it is bigger.

A test experiment based on a FSM is modelled by the notion of  \emph{checking
sequence}, i. e. a finite sequence of inputs that distinguishes by some output  the
specification FSM from any other FSM with at most the same number of states.
\begin{definition}
Let $A$ be a specification FSM with $n$ states and initial state $s_0$. 
A {\em checking sequence} for $A$ is an input sequence $\sigma_{check}$ such that
for every FSM $B$ with initial state $s'_0$, the same input alphabet,  and at most
$n$ states, that is not isomorphic to $A$,
   $\lambda^*_B(s'_0, \sigma_{check}) \neq \lambda^*_A(s_0,  \sigma_{check})$.
\end{definition}

The complexity of the construction of checking sequences depends on two important
characteristics of the specification FSM: the existence of a {\em reliable reset}
that makes it possible to start the test experiment from a known state, and the
existence of a {\em distinguishing sequence} $\sigma$, which can identify the
resulting state after an input sequence, i.e.  such that for every pair of distinct
states $s_i$, $s_j$, 
$\lambda^* (s_i, \sigma) \neq \lambda^*(s_j,\sigma)$.

A reliable reset is a specific input symbol that leads an FSM from any state to the
same state: for every state $s$, $\delta(s, reset) = s_r$.
For FSM without reliable reset, the so-called {\em homing sequences} are used to
start the checking sequence.
A {\em homing sequence} is an input sequence $\sigma_h$ such that, from any state,
the output sequence produced by $\sigma_h$ determines uniquely the arrival state.
For every pair of distinct states 
$s_i, s_j, \lambda^*(s_i, \sigma_h) = \lambda^*(s_j,\sigma_h)$ implies
$\delta^*(s_i, \sigma_h) = \delta^*(s_j,\sigma_h)$.
Every reduced FSM has an homing sequence of polynomial length, constructible in
polynomial time.

The decision whether the behavior of the IUT is satisfactory, requires to observe 
the states of the IUT  either before or after some action. 
As the IUT is a running black box system, the only means of observation is by
submitting other inputs and collecting the resulting outputs. 
Such observations are generally destructive as they may change the observed state.

The existence of a distinguishing sequence makes the construction of a
checking sequence  easier:
an example of a checking sequence for a FSM $A$ is a sequence of inputs resulting in
a trace that traverses  once every
transition followed by this distinguishing sequence to detect for every transition
both output errors and errors of arrival state.

Unfortunately deciding whether a given FSM has a distinguishing sequence is
PSPACE-complete with respect to the size of the FSM (i.e. the number of states).
However, it is polynomial for adaptative distinguishing sequences (i.e input trees
where choices of the next input are guided by the outputs of the IUT), and it is
possible to construct one of quadratic length.
For several variants of these notions, see \cite{LY}.

Let $p$ the size of the input alphabet.
For an FSM with a reliable reset, there is a polynomial time algorithm, in
$O(p.n^3)$, for constructing a checking sequence of polynomial length, also in
$O(p.n^3)$ \cite{Vasilevskii1973,Chow1978}.
For an  FSM with a distinguishing sequence there is a deterministic polynomial time
algorithm to construct a checking sequence \cite{Hennie,Kohavi} of length polynomial
in the length of the distinguishing sequence.

In other cases, checking sequences of polynomial length also exist,
but finding them requires more involved techniques such as randomized  
algorithms.
More precisely, a randomized  algorithm can construct with high probability in  
polynomial time a checking sequence of length $O(p.n^3 + p'.n^4. \log n)$,
with $p' = min(p, n)$.
The only known deterministic complexity of producing such 
 sequences is exponential either in time or in the length of the checking sequence.

The above definitions and results generalize to the case where FSM $B$ has more
states than FSM $A$. The complexity of generating checking sequences, and their
lengths, are exponential in the number
of extra states.

\subsubsection{Non determinism} \label{nondet}

The concepts presented so far are suitable when both the specification FSM and the
IUT are deterministic. 
Depending on the context and of the authors, a non deterministic specification FSM
$A$ can have different meanings: it may be understood as describing a class of
acceptable deterministic implementations or it can be understood as describing some
non deterministic acceptable implementations.
In both cases, the notion of equivalence of the specification FSM $A$ and of the
implementation FSM $B$ is no more an adequate basis for testing.
Depending of the authors, the required relation between a specification and an
implementation is called the ``satisfaction relation'' ($B$ satisfies $A$) or the
``conformance relation'' ($B$ conforms to $A$).
Generally it is not an equivalence, but a preorder (see \cite{Tretmans92,GJ98,BT}
among many others).

A natural definition for this relation could be the so-called ``trace inclusion''
relation: 
any trace of the implementation must be a trace of the specification. 
Unfortunately, this definition accepts, as a conforming implementation of any
specification, the idle implementation, with an empty set of traces. 
Several more elaborated relations have been proposed. 
The most known are the \emph{conf} relation, between Labelled Transition Systems
\cite{Brinksma} and the \emph{ioco} relation for Input-Output Transition Systems
\cite{Tretmans96}.
The intuition behind these relations is that when a trace $\sigma$ (including the
empty one) of a specification $A$ is executable by some IUT $B$, after $\sigma$, $B$
can be idle only if $A$ may be idle after $ \sigma $, 
else $B$ must perform some action performable by $A$ after $\sigma$. 
For Finite State Machines, it can be rephrased as: an implementation FSM $B$
conforms to a specification FSM $A$ if all its possible responses to any input
sequence could have been produced by $A$, a response being the production of an
output or idleness.

Not surprisingly, non determinism introduces major complications when testing. 
Checking sequences are no more adequate since some traces of the specification FSM
may not be executable by the IUT. One has to define \emph{adaptative} checking
sequences (which, actually, are covering trees of the specification FSM) in order to
let the IUT choose non-deterministically among the allowed behaviors.

\subsubsection{Symbolic traces and constraints solvers}\label{efsm}

Finite state machines (or finite transition systems) have a limited description power. 
In order to address the description of realistic systems, various notions of
Extended Finite State Machines (EFSM) or symbolic labelled transition systems (SLTS)
are used.
They are the underlying semantic models in a number of industrially significant
specification techniques, such as LOTOS, SDL, Statecharts, to name just a few. 
To make a long story short, such models are enriched by a set of typed variables
that are associated with the states.
Transitions are labelled as in FSM or LTS, but in addition, they have associated
guards and actions, that are conditions and assignments on the variables.
In presence of such models, the notion of a checking sequence is no more realistic. 
Most EFSM-based testing methods derive some test set from the EFSM, that is a set of
input sequences that ensure some coverage of the EFSM, assuming some uniform
behavior of the IUT with respect
to the conditions that occur in the EFSM.

More precisely, an {\em Extended Finite State Machine} (EFSM) is a  structure
$(S,s_0,I,IP,O,T,V, \vec{~v_0})$ where $S$ is a finite set of states with initial
state $s_0$, $I$ is a set of input values and $IP$ is a set of input parameters
(variables), $O$ is a set of output values, $T$ is a finite set of symbolic
transitions, $V$ is a finite list of variables and
$\vec{~v_0}$ is a list of initial values of the variables.
Each association of a state and variable values is called a configuration.
Each symbolic transition $t$ in  $T$ is a 6-tuple: $t = (s_t , s'_t , i_t , o_t ,
G_t , A_t )$ where 
$s_t , s'_t $  are respectively the current state, and the next state of $t$; 
$i_t$ is an input value or an input parameter;
$o_t$ is an output expression that can be parametrized by the variables and the
input parameter. 
$G_t $ is a predicate (guard) on the current variable values and the input parameter
and $A_t $ is an update action on the variables that may use values of the variables
and of the input.
Initially, the machine is in an initial state $s_0$ with initial variable values:
$\vec{~v_0}$.

\begin{figure}[h] 
\centerline{\includegraphics[width=15cm]{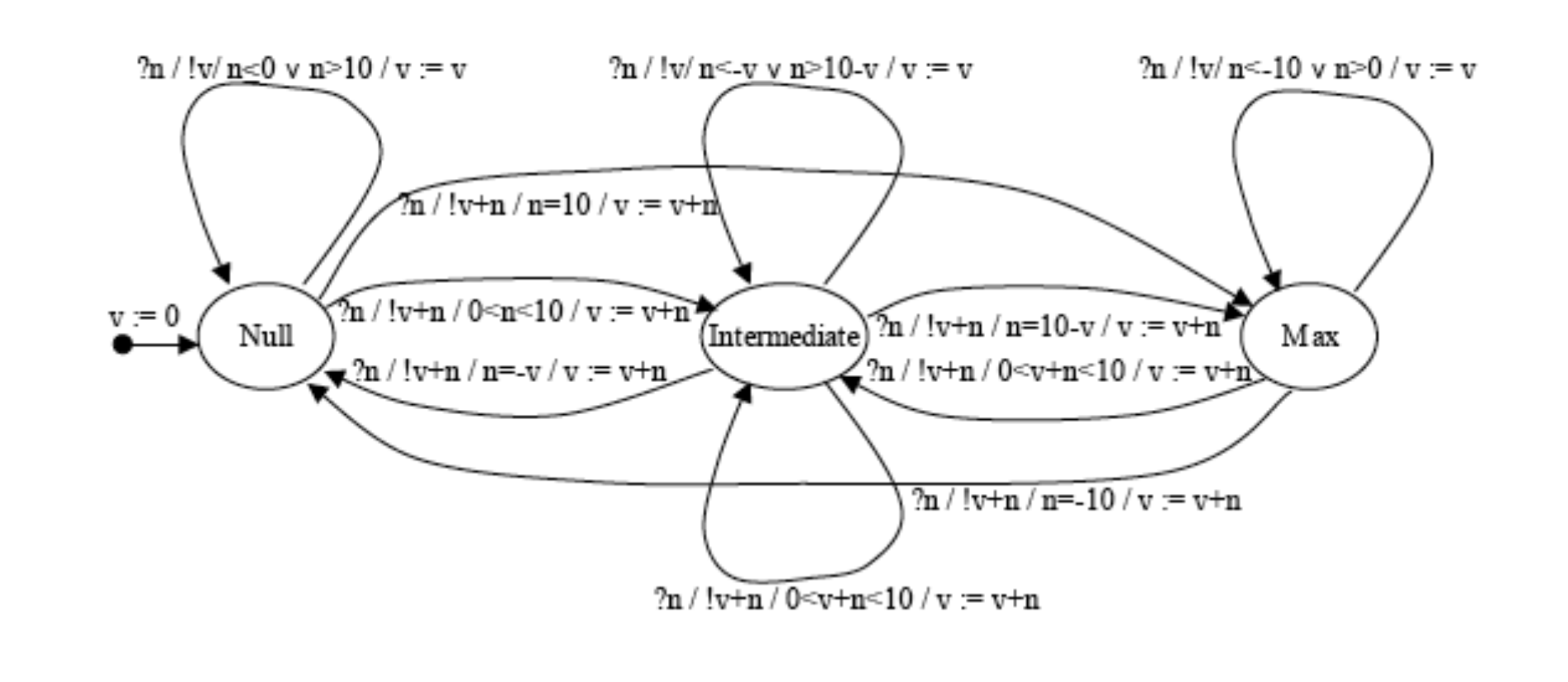}}
\caption{Example of an EFSM: counter with increment and decrement values. }\label{EFSM}
\end{figure}

An action  $v :=v+n$  indicates the update of the variable $v$. 
Figure \ref{EFSM} gives a very simple example of such an EFSM. It is a bounded counter which  receives increment or decrement values. 
There is one state variable $v$ whose domain is the integer interval 
$ [0 ..10 ]$. 
 The variable $v$ is initialized to $0$. The input domain $I$ is $\mathcal{Z} $. There is one integer input parameter $n$. When an input would provoke an overflow or an underflow of $v$, it is ignored and $v$ is unchanged.
Transitions labels follows the following syntax:
$$
? <input \ value \  or \ parameter>/! <output \ expression>/<guard>/<action> $$

An EFSM operates as follows: in some configuration, it receives some input and computes
the guards that are satisfied for the current configuration.
The satisfied guards identify enabled transitions. 
A single transition among those enabled is fired. 
When executing the chosen transition, the EFSM 
\begin{itemize}
\item reads the input value or parameter value $i_t$,
\item updates the variables according to the action of the transition,
\item moves from the initial to the final
state of the transition,
\item produces some output , which is
computed from the values of the variables and of the input
via the output  expression of the transition. 
\end{itemize}

Transitions are atomic and cannot be interrupted.
Given an EFSM, if each variable and input parameter has a finite number of values
(variables for booleans or for intervals of finite integers,  for example), then
there is a finite number of configurations, and hence there is a large  equivalent
(ordinary) FSM with configurations as states. 
Therefore, an EFSM with finite variable domains is a succinct representation of an FSM.
Generally, constructing this FSM is not easy because of the reachability problem,
i.e. the issue of determining if a configuration is reachable from the initial
state. It is undecidable if the variable domains are infinite and PSPACE-complete
otherwise\footnote{As said above, there are numerous variants of the notions of EFSM
and SLTS. The complexity of their analysis (and thus of their use as a basis for
black box testing) is strongly dependent on the types of the variables and of the
logic used for the guards.}.

A {\em symbolic trace} $t_1,\ldots, t_n$ of an EFSM is a sequence of symbolic
transitions such that $s_{t_1} = s_0$ and for $i = 1, \ldots n-1$, $s'_{t_i} =
s_{t_{i+1}}$.
A {\em trace predicate} is the condition on inputs which ensures
the execution of a symbolic trace.
Such a predicate is built by traversing the trace $t_1,\ldots, t_n$ 
in the following way:
 \begin{itemize}
  \item the initial index of each variable $x$ is $0$, and for each variable $x$
there is an equation $x_0 = v_0$,
  \item for $ i = 1 \ldots n$, given transition $t_i$ with guard $G_i$, and action
$A_i$:
  \begin{itemize}
   \item guard $G_i$ is 
transformed into the formula $\tilde G_i$  where each variable of $G$ has been 
indexed by its current index, and the input parameter (if any) is indexed by $i$,
  \item each assignment in $A_i$ of an expression $\, expr$ to some
  variable $x$ is transformed into an 
equation $x_{k+1} = \widetilde{expr}_i$ where $k$ is the current index of  
$x$ and $\widetilde{expr}_i$ is the expression $expr$ where each 
variable is indexed by its current index, and the input parameter (if any) is
indexed by $i$,
\item the current indexes of all assigned variables are incremented, 
 \end{itemize}
\item the trace predicate is the conjunction of all these formulae.
\end{itemize}

A symbolic trace is {\em feasible} if its predicate is satisfiable, i.e. there exist
some sequence of input values that ensure that at each step of the trace, the guard
of the symbolic transition is true. 
Such a sequence of inputs characterizes a trace of the EFSM. 
A configuration is reachable if there exists a trace leading to it.

EFSM testing methods must perform reachability analysis: to compute some input
sequence that exercises a feature (trace, transition, state) of a given EFSM, a
feasible symbolic trace leading to and covering this feature must be identified and
its predicate must be solved. 
Depending on the kind of formula and expression allowed in guards and actions,
different constraint solvers may be used \cite{C11,T11}. 
Some tools combine them with SAT-solvers, model checking techniques, symbolic
evaluation methods including abstract interpretation, to eliminate some classes of
clearly infeasible symbolic traces.

The notion of EFSM is very generic. The corresponding test generation problem is
very similar to test generation for programs in general.
The current methods address specific kinds of EFSM or SLTS. There are still a lot of
open problems to improve the levels of generality and automation.

\subsubsection{Classical methods  in probabilistic and statistical testing}\label{pst}
 
 Drawing test cases at random is an old idea, which looks attractive at first sight.
 It turns out that it is difficult to estimate its detection power. Strong
hypotheses on the IUT, on the types and distribution of faults, are necessary to
draw conclusions from such test campaigns.
 Depending on authors and contexts, testing methods based on random selection of
test cases are called: random testing, or probabilistic testing or statistical
testing.
 These methods can be classified into three categories : 
those based on the input domain,  
those based on the environment,
and those based on some knowledge of the behavior of the IUT.

In the first case, classical random testing (as studied  in \cite{DuNt81,DuNt84})
consists in selecting
test data uniformly at random from the input domain of the program.
In some variants, some knowledge on the input domain is exploited, for instance to
focus on the boundary or limit conditions of the software being tested
\cite{Reid97,Nt01}.

In the second case, the selection is based on an operational profile, i.e. 
an estimate of the relative frequency  of inputs.  
Such testing methods are called {\em statistical testing}. 
They can serve as a statistical sampling method to collect failure data for
reliability estimation
(for a survey see \cite{MuFuIr96}).

In the third case, some description of the behavior of the IUT is used.
In  \cite{ThWa91},  the choice
of the distribution on the input domain is guided either by some coverage
criteria of the program and they call their method {\em structural statistical
testing}, or by some
specification and they call their method {\em functional statistical testing}. 

Another approach is to perform random walks \cite{Ald91} 
in the set
of execution paths or traces of the IUT.
Such testing methods  were developed early in the area
of communication protocols  \cite{West89,MP94}. 
In \cite{West89}, West reports experiments where random walk
methods had good and stable error detection power.
In \cite{MP94},  some class of models is identified, namely
those where the underlying graph is symmetric, 
which can be efficiently tested by random walk exploration: under this strong
condition, the
random walk converges to the uniform distribution over the state space
in polynomial time with respect to the size of the model.
 A general problem with all these methods is the impossibility, except for some
  very special cases, to assess the results of a test campaign, either in term of
coverage or in term of fault detection.

\section{Methods for approximation}
In this section we classify the different  approximations
introduced  in 
model checking and testing in two categories. Methods which approximate decision problems, 
based on some parameters, and methods which study {\em approximate versions of the decision problems}.
\begin{enumerate}

\item Approximate methods for decision, counting and learning problems. 
The goal is to define useful heuristics on practical inputs. 
SAT is the typical example where no polynomial algorithm exists assuming $P\neq NP$, 
but where useful heuristics are known. 
The search for abstraction methods by successive refinements follows the same approach.
 
\item   Approximate versions of decision and learning problems relax the decision by introducing some error parameter $\eps$. In this case, we may obtain efficient randomized algorithms, often based on statistics for these new approximate decision problems.  
\end{enumerate}

Each category is detailed in subsections below. First, we introduce  the classes of efficient algorithms we will use to elaborate approximation methods.

\subsection{Randomized algorithms and complexity classes}

The efficient algorithms we study are mostly randomized algorithms which operate in polynomial time.
They use an extra instruction,  {\em flip a coin}, and we
obtain $0$ or $1$ with probability $\frac{1}{2}$. 
As we make $n$ random flips,
the probabilistic space $\Omega$ consists of all binary sequences of length $n$,
each with probability $\frac{1}{2^n}$. 
We want to decide if $x\in L \subseteq \Sigma ^*$, such that the probability of getting  the wrong answer is less
than $\frac{c}{2^n}$ for some fixed constant $c$, i.e. exponentially small.

\begin{definition} An algorithm  $\mathcal{A}$ is   {\em Bounded-error
Probabilistic Polynomial-time} (BPP), for a language $L \subseteq \Sigma ^*$ if
$\mathcal{A}$ is in polynomial time and:
  
\begin{itemize}
\item if $x \in  L$ then $\mathcal{A}$ accepts $x$ with probability greater then $2/3$,
\item if $x \not\in  L$ then $\mathcal{A}$ rejects $x$ with probability greater then
$2/3$.
\end{itemize}
The class BPP
consists of all languages $L$ which admit a bounded-error
probabilistic polynomial time algorithm.
\end{definition}

In this definition, we can replace $2/3$ by any value strictly greater than $1/2$,
and obtain an equivalent definition. In some cases, $2/3$ is replaced by $1/2+\eps$
or by 
$1- \delta$ or by $1-1/n^k$. If we modify the second condition of the previous defintion by: 
if $x \not\in L$ then $\mathcal{A}$ rejects $x$ with probability 1, we obtain the
class RP, {\em Randomized Polynomial time}.

We recall the notion of a p-predicate, used to define the class $\NP$ of decision problems which are verifiable in polynomial time.

\begin{definition} A {\bf p-predicate} $R$ is a binary
relation between words such that there exist two polynomials
$p , q$ such that:
\begin{itemize} 
\item for all ${\bf x},{\bf y} \in \Sigma ^* $, 
$R({\bf x},{\bf y}) $ implies that $\mid{\bf y} \mid \leq p (\mid{\bf
x} \mid) $; 
\item for all ${\bf x},{\bf y} \in \Sigma ^* $ , $R ({\bf
x},{\bf y}) $ is decidable in time $q (\mid{\bf x} \mid) $.
\end{itemize} 
\end{definition}

A decision problem $A$ is in the class $\NP$ if there is a p-predicate $R$ such that for all $x$, $x\in A$ iff $\exists y R(x,y)$. Typical examples are $\SAT$  for clauses
or $\CLIQUE$ for graphs. 
For $\SAT$, the input $x$ is a set of clauses, $y$ is a valuation and $R(x,y)$ if
$y$ satisfies $x$.  For  $\CLIQUE_k$, the input $x$ is a graph, $y$ is a subset of size $k$ of the
nodes and  $R(x,y)$ if $y$ is a clique of $x$, i.e. if all pairs of nodes in $y$ are
connected by an edge. 

One needs a precise notion of approximation for a counting function
 $F:\Sigma ^ * 
\rightarrow N$ using an efficient randomized algorithm whose  relative error 
is bounded by  $\epsilon$  with  high 
probability, for all  $\epsilon$. It is used in section \ref{sras} to
approximate probabilities.

 \begin{definition}\label{fpras} An algorithm $\mathcal{A}$ is a {\em
Polynomial-time 
Randomized Approximation Scheme} (PRAS)
  for a function $F:\Sigma ^ * \rightarrow \mathbb{N}$ if for 
every $\epsilon$ and  {\bf x},  
\[\B \{\mathcal{A}({\bf x}, \epsilon) \in 
[(1-\epsilon). F ({\bf x}) , (1 +\epsilon). F ({\bf x})] \} ~ \geq ~ \frac{2} 
{3} \] and $\mathcal{A} ({\bf x}, \epsilon) $ stops in  polynomial time 
in $\mid {\bf x} \mid$. The algorithm $\mathcal{A}$ is a {\em Fully Polynomial time Randomized Approximation 
Schema} (FPRAS),  if the time 
of computation is also polynomial in  $1/\epsilon $. The class PRAS (resp. FPRAS)
consists of all functions  $F$ which admits a PRAS (resp. FPRAS) .
\end{definition} 
 
If the algorithm $\mathcal{A}$ is deterministic, one speaks of an $PAS$ and of a 
$FPAS$. A $PRAS (\delta) $ (resp. $FPRAS ( \delta ) $), is an algorithm 
 $\mathcal{A}$ which outputs a value $\mathcal{A}({\bf x}, \epsilon 
, \delta) $  such that: 
 \[\B \{ \mathcal{A}({\bf x}, \epsilon , 
\delta) \in [(1-\epsilon). F ({\bf x}), (1 +\epsilon). F ({\bf x})] \} ~ \geq 
~ 1 - \delta \] and whose time complexity is also polynomial in $\log 
(1 / \delta) $. The error probability is less than $\delta$ in this 
model. In general, 
 the
probability of success can be amplified 
from $2/3$ to $1-\delta$ at the cost of extra  
computation  of length polynomial in $\log ( 1 / \delta ) $. 
 
\begin{definition} A counting function $F$ 
is in the class  $\#$P if there exists a p-predicate $R$ such that for all $x$, 
 $F(x)=\mid\{y : (x,y) \in R\}\mid$. 
 \end{definition} 
 If $A$ is an $\NP$ problem, i.e. the decision
problem on input $x$ which decides if there exists $y$ such that 
$R(x,y)$ for a p-predicate $R$, then $\#A$ is the associated counting function, i.e.
$\#A(x)=\mid\{y : (x,y) \in R\}\mid$.
 The counting  problem 
$\#SAT$ is $\#$P-complete and 
not approximable (modulo some complexity conjecture). On the other hand
  $\#DNF$ is also $\#$P-complete but admits an $FPRAS$ \cite{pa:KaLu}.

\subsection{Approximate methods for satisfiability, equivalence, counting and learning}

{\em Satisfiability } decides given a model  $\mathcal{M}$ and a formula $\psi$, whether  $\mathcal{M}$  
satisfies a formula $\psi$. {\em Equivalence} decides given two models $\mathcal{M}$ and $\mathcal{M}'$, 
whether they satisfy the same class of formulas. {\em Counting} associates to a formula $\psi$, the number of models 
 $\mathcal{M}$ which  satisfy a formula $\psi$. {\em Learning } takes a black box which defines an unknown function $f$
 and  tries to find from samples $x_i, y_i=f(x_i)$.

 \subsubsection{Approximate satisfiability and abstraction}\label{aia}

To  verify that a model $\mathcal{M}$ satisfies a formula $\psi$ , abstraction can
be used for constructing approximations of $\mathcal{M}$ that are sufficient for
checking $\psi$.
This approach goes back to the notion of \textit{Abstract Interpretation}, a theory
of semantic approximation 
of programs introduced by Cousot et al.\cite{cc77}, which constructs elementary
embeddings\footnote{Let $U$ and $V$ be two structures with domain $A$ and $B$.
In logic, an {\em elementary embedding} of  $U$ into $V$ is a function
$f: A \rightarrow B$ such that for all formulas $\varphi(x_1,...,x_n)$ of a logic,
for all elements $a_1,...,a_n \in A$,  $U\models \varphi[a_1,...,a_n] ~ {\rm iff}~ 
V\models \varphi[f(a_1),...,f(a_n)]$.}
that suffice to decide properties of programs. A classical example is
multiplication, where modular arithmetic is the basis of the abstraction. It has
been applied in static analysis to find  sound, finite, and approximate
representations of a program.

 In the framework of model checking, 
 reduction by \textit{abstraction} consists in approximating infinite or very large 
finite transition systems by finite ones, on which existing algorithms designed for 
finite verification are directly applicable. 
This idea was first introduced by Clarke et al. \cite{cgl94}. 
Graf and Saidi  \cite{gs97} have then proposed the {\em predicate abstraction} 
method where abstractions are automatically obtained, using decision procedures,
from a set of predicates given by the user. 
When the resulting abstraction is not adequate for checking $\psi$, the set of
predicates must be revised.
This approach by \textit{abstraction refinement} has been recently systematized,
leading to a quasi
automatic  abstraction discovery method known as {\em Counterexample-Guided Abstraction
Refinement} (CEGAR) \cite{cegar}. It relies on the iteration of three kinds of steps:
abstraction construction,
model checking of the abstract model, abstraction refinement, which, when it
terminates, states whether the original model satifies the formula.

This section starts with the  notion of abstraction used in model checking, based on
the pioneering paper by Clarke et al..
Then, we present the principles of predicate abstraction and abstraction refinement.

In \cite{cgl94}, Clarke and al. consider transition systems $\mathcal{M}$ where
atomic propositions are formulas of the form $v=d$,
where $v$ is a variable and $d$ is a constant.
Given a set of typed variable declarations $v_1:T_1, \ldots , v_n:T_n$,  
states can be
identified with n-tuples of values for variables, and the labeling function $L$ is
just defined
by $L(s)=\{s\}$.
On such systems, abstractions can be defined by a surjection for each variable into
a smaller domain. It reduces the size of the set of states. Transitions are then
stated between the resulting equivalence classes of states as defined below.

\begin{definition}(\cite{cgl94})
Let $\mathcal{M}$ be a transition system, with set of states $S$, transition
relation $R$, and a set of initial states
$I \subseteq S$.
An {\em abstraction for $\mathcal{M}$} is 
a surjection $h:S\rightarrow \widehat{S}$.
A transition system 
$\widehat{\mathcal{M}}= ( \widehat{S},\widehat{I},\widehat{R}, \widehat{L} )$
{\em approximates $\mathcal{M}$ with respect to $h$} 
($\mathcal{M}\sqsubseteq_h\widehat{\mathcal{M}}$ for
short) if  
$h(I) \subseteq 
\widehat{I}$
and $(h(s),h(s'))\in\widehat{R}$ for all $(s,s')\in R$.
\end{definition}
Such an approximation is called an {\em over approximation} and is explicitly given in
\cite{cgl94} from a given logical representation of $\mathcal{M}$.

Now, let $\widehat{\mathcal{M}}$ be an approximation of $\mathcal{M}$.
Suppose that $\widehat{\mathcal{M}}\models \Theta$. What can we conclude
on the concrete model $\mathcal{M}$?
First consider the following transformations $\mathcal{C}$
and 
$\mathcal{D}$ between $\mathrm{CTL}^{*}$
formulas on $\mathcal{M}$ and their approximation on 
$\widehat{\mathcal{M}}$. 
These transformations 
preserve boolean 
connectives, path quantifiers, and temporal operators,
and act on atomic propositions as follows:
$$\mathcal{C}(\widehat{v}=\widehat{d})\eqdef
\bigvee_{d : h(d)=\widehat{d}} 
(v=d),\qquad
\mathcal{D}(v=d)\eqdef
(\widehat{v}=h(d)).$$

Denote by $\boldsymbol{\forall}\mathrm{CTL}^{*}$ and 
$\boldsymbol{\exists}\mathrm{CTL}^{*}$ the universal fragment and 
the existential fragment of $\mathrm{CTL}^{*}$.
The following theorem gives
correspondences between models and their approximations.

\begin{theorem}[\cite{cgl94}]\label{transfer}
Let $\mathcal{M}= (S, I, R, L)$ be a transition system.
Let $h:S\rightarrow\widehat{S}$ be an abstraction for $\mathcal{M}$, and let
$\widehat{\mathcal{M}}$ be such that $\mathcal{M}\sqsubseteq_h\widehat{\mathcal{M}}$.
Let $\Theta$ be a $\boldsymbol{\forall}\mathrm{CTL}^{*}$ formula on $\widehat{M}$,
and $\Theta'$ be a $\boldsymbol{\exists}\mathrm{CTL}^{*}$ formula on $M$.
Then 
$$\widehat{\mathcal{M}}\models\Theta \implies \mathcal{M}\models\mathcal{C}(\Theta)
\quad \text{and} \quad
\mathcal{M}\models\Theta' 
\implies \widehat{\mathcal{M}}\models\mathcal{D}(\Theta').$$
\end{theorem}

Abstraction can also be used when the target structure does not follow the original
source signature. In this case, some specific new predicates define the target
structure and the technique has been called {\em predicate abstraction} by Graf et
al. \cite{gs97}. The analysis of the small abstract structure may suffice to prove a
property of the concrete model and the authors
 define a method  to construct {\em abstract state
graphs}
from models of concurrent processes with variables on finite domains. In these
models, transitions are labelled  by guards and assignments. The method starts from
a given set of predicates on the variables. The choice of these predicates is
manual, inspired by the guards and assignments occurring on the transitions. The
chosen predicates induce equivalence classes on the states. The computation of the
successors of an abstract state requires theorem proving. Due to the number of
proofs to be performed, only relatively small abstract graphs can be constructed. As
a consequence, the corresponding approximations are often rather coarse. They must
be tuned, taking into account the properties to be checked.

\begin{figure}[h]
\centerline{\includegraphics[height=5cm]{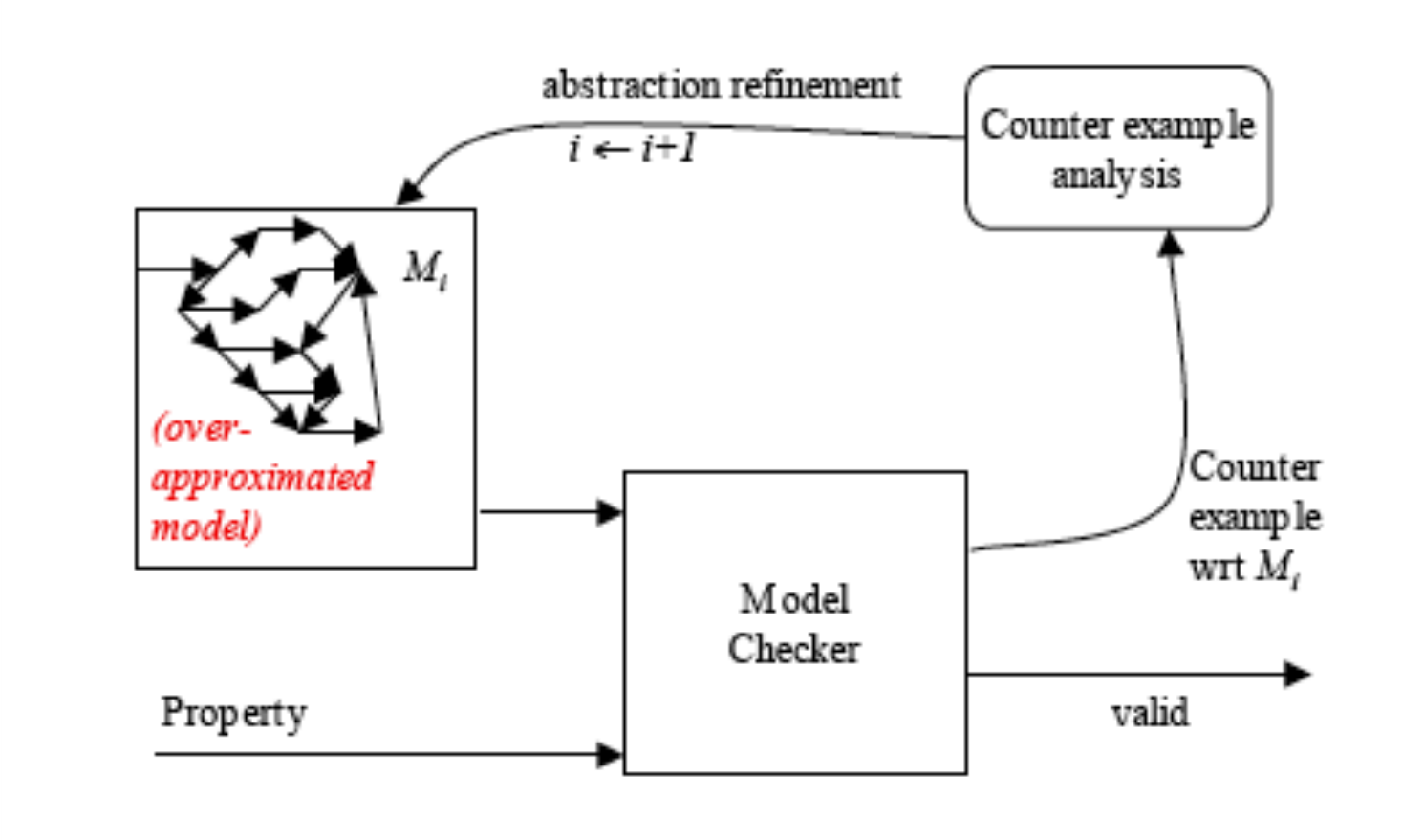}}
\caption{CEGAR:Counterexample-Guided Abstraction Refinement. }\label{cegardraw}
\end{figure}

We now explain how to use abstraction refinement in order to achieve $\forall CTL^*$
model checking: for a concrete structure $\mathcal{M}$ and an $\forall CTL^*$
formula $\psi$, we would like to check if $M \models \psi$. The methodology of the
counterexample-guided abstraction refinement  \cite{cegar} consists in the
following steps:
\begin{itemize}
\item Generate an initial abstraction $\widehat{\mathcal{M}}$.
\item Model check the abstract structure. If the check is affirmative, one can
conclude that $\mathcal{M} \models \psi$; otherwise, there is a counterexample to
$\widehat{\mathcal{M}} \models \psi$. To verify if it is a real counterexample, one
can check it on the original structure; if the answer is positive, it is reported it
to the user; if not, one proceeds to the refinement step.
\item Refine the abstraction by partitioning equivalence classes of states so that
after the refinement, the new abstract structure does not admit the previous
counterexample. After refining the abstract structure, one returns to the model
checking step.
\end{itemize}

The above approaches are said to use {\em over approximation} because  the reduction
induced on the models introduces new paths, while preserving the original ones.
A notion of {\em under approximation} is used in bounded model checking where paths
are restricted to some finite lengths. It is presented in section \ref{bmc}. 
Another approach using under approximation is taken in \cite{ms07} for the class of
models with input
variables. The original model is coupled with a well chosen logical circuit with $m
< n$ input variables and
$n$ outputs. The model checking of the new model may be easier than the original
model checking, as fewer input variables are considered.

\subsubsection{Uniform generation and counting}
\label{counting}

In this section we describe the link between generating elements of a set $S$ and
counting the size of $S$, first in the exact case and then in the
approximate case. The exact case is used in section \ref{crt} and the
approximate case is later used in section \ref{sras} to approximate
probabilities.\\

{\bf Exact case.}
Let $S_n$ be a set of combinatorial objects of size $n$.
There is a close connection between having an explicit formula for $\mid S_n
\mid$ and  a uniform generator for objects in $S_n$.
Two major approaches have been developed for counting and drawing uniformly at
random combinatorial structures: 
the Markov Chain Monte-Carlo approach (see e.g. the survey \cite{JS96}) and the
so-called recursive method, as described in \cite{Fla} and implemented in
\cite{mupad}. 
Although the former is more general in its applications, the latter is particularly
efficient for dealing with the so-called {\em decomposable
  combinatorial classes of Structures}, namely classes where
structures are formed  from a set $ \mathcal{Z}$ of given
\emph{atoms} combined by the following constructions:
$$
 +, \times, \textsc{Seq}, \textsc{PSet}, \textsc{MSet}, \textsc{Cyc}
$$
respectively corresponding to disjoint union, Cartesian product, finite sequence,
multiset, set, directed cycles. It is possible to state cardinality constraints via
subscripts (for instance $ \textsc{Seq}_{\le 3}$).
These structures are called \emph{decomposable structures}.
The size of an object is the number of atoms it contains.

\begin{example}
Trees : 
\begin{itemize}
\item The class $ \mathcal{B} $ of   binary trees can be specified by the equation
$ \mathcal{B} = \mathcal{Z} + (\mathcal{B} \times \mathcal{B})$
where $\mathcal{Z}$ denotes a fixed set  of atoms. 
\item An example of a structure  in $ \mathcal{B}$ is $(\mathcal{Z} \times (\mathcal{Z}
\times \mathcal{Z}))$. Its size is 3.
\item For non empty ternary trees one could write 
$ \mathcal{T} = \mathcal{Z} + \textsc{Seq}_{= 3}(\mathcal{T})$
\end{itemize}
\end{example}

The enumeration of decomposable structures is based on generating functions. Let
$C_n$ the number of objects of $C$ of size $n$, and the following generating
function:
$$
C(z) = \sum_{n \le 0} C_n z^n
$$
Decomposable structures can be translated into generating functions using classical
results of combinatorial analysis. A comprehensive dictionary
is given in \cite{Fla}.
The main result on counting and random generation of decomposable structures is:

\begin{theorem}
Let $C$ be a  decomposable combinatorial class of structures. Then the counts $\{ C_j |
j =0 \ldots n   \}$ can be computed in $O(n^{1+\epsilon})$ arithmetic
operations, where $\eps$ is a constant less than $1$. 
In addition, it is possible to draw an element of size $n$ uniformly at random in
$O(n \log n)$ arithmetic operations in the worst case.
\end{theorem}
A first version of this theorem, with a computation of the counting sequence $\{ C_j
| j =0 \ldots n   \}$ in $O(n^2)$ was given in \cite{Fla}. The improvement to 
$O(n^{1+\epsilon})$ is due to van der Hoeven \cite{VHoe2002}. 

This theory has led to powerful practical tools for random generation \cite{mupad}.
There is a preprocessing step for the construction of the  $\{ C_j | j =0 \ldots n  
\}$ tables . Then the drawing is performed following the decomposition pattern of
$C$, taking into account the cardinalities of the involved sub-structures.
For instance, in the case of binary trees, one can uniformly generate binary trees
of size $n+1$
 by generating a random $k\leq n$, with probability
$$p(k) = \frac{|\mathcal{B}_k|. |\mathcal{B}_{n-k}|} { | \mathcal{B}_n|}$$ 
where  $\mathcal{B}_k$ is the set of binary trees of size $k$. A tree
of size $n+1$ is decomposed into a subtree on the
left side of the root  of size $k$ and into a  subtree on the right side of
the root  of size $n-k$.
One recursively applies this procedure and generates a binary tree with $n$
atoms following a uniform distribution on $\mathcal{B}_n$.\\

{\bf Approximate case.}
In the case of a hard counting problem, i.e. when $\mid S_n \mid$ does not have an
explicit formula, one can introduce a useful approximate version of counting and
uniform generation. Suppose the  objects are witnesses of a p-predicate, i.e. they
can be recognized in polynomial time.

Approximate counting $S$ can be reduced to approximate uniform generation of $y\in
S$ and conversely 
approximate uniform generation can be reduced to approximate counting, for
self-reducible sets. Self-reducible sets guarantees that a solution for an instance of size $n$
depends directly from solutions for instances of size $n-1$. For
example, in the case of SAT, a
valuation on $n$ variables $p_1,...,p_n$
 on an instance $x$ 
is either a valuation of an instance $x_1$ of size  $n-1$ where $p_n=1$ or a
valuation of an instance $x_0$ of size  $n-1$ where $p_n=0$.
Thus the p-predicate for SAT is a self-reducible relation.

To reduce approximate counting to approximate uniform generation,
let $S_{\sigma}$ be the set
$S$ where the first letter of $y$ is $\sigma$, and
$p_{\sigma}=\frac{\mid S_{\sigma}\mid}{\mid S \mid}$. For self-reducible sets 
 $\mid S_{\sigma}\mid$  can be recursively approximated using the same technique.
 Let $p_{\sigma.\sigma'}=\frac{\mid S_{\sigma.\sigma'}\mid}
{\mid S_{\sigma} \mid}$ and so on,  until  one reaches
$\mid S_{\sigma_1,...,\sigma_m}\mid$ if $m=|y|-1$, which can be directly computed. 
Then $$\mid S \mid=\frac{\mid S_{\sigma_1,...,\sigma_m}\mid}
{p_{\sigma_1}.p_{\sigma_1.\sigma_2},...,p_{\sigma_1,...,\sigma_{m-1}}}$$
Let
$\widehat{p_{\sigma}}$ be the estimated measure for $p_{\sigma}$ obtained with the
uniform generator for $y$. The $p_{\sigma_1,...,\sigma_i}$ can be replaced by their
estimates
 and leading to an estimator for $\mid S \mid$.

Conversely, one can reduce approximate uniform generation to approximate counting.
Compute $\mid S_{\sigma}\mid$ and $\mid S \mid$. Suppose 
$\Sigma=\{0,1\}$ and let $p_{0}=\frac{\mid S_{0}\mid}{\mid S \mid}$. 
Generate $0$ with probability $p_0$ and $1$ with probability
$1-p_0$ and recursively apply the same method. If one obtains $0$ as the first bit,
one sets 
$p_{00}=\frac{\mid S_{00}\mid}{\mid S_0 \mid}$ and generates $0$ as the next bit
with  probability $p_{00}$ and $1$ with probability
$1-p_{00}$, and so on.
One obtains a string $y \in S$ 
with an approximate uniform distribution.

\subsubsection{Learning}

In the general setting, given a black box, i.e.
an unknown function $f$, and samples $x_i, y_i=f(x_i)$ for $i=1,...,N$, one wishes to find $f$.
Classical learning theory distinguishes between supervised and unsupervised learning.
In supervised learning $f$ is one function among a class ${\cal F} $
of given functions. In unsupervised learning, one tries to find $g$ as the best possible function.

Learning models suppose {\em membership queries}, i.e. positive and negative
examples, i.e.
 given $x$, an oracle produces $f(x)$ in one step.
Some models assume more general queries such as  {\em conjecture queries}:
given an hypothesis $g$, an {\em oracle} answers YES if $f=g$, else produces
an $x$ where $f$ and $g$ differ.
For example, let 
 $f$ be a function  $\Sigma^*  \rightarrow \{0,1\}$ 
where  $\Sigma$ is a finite alphabet. It describes a
language  $L= \{x \in \Sigma^*, ~~f(x)=1\} \subseteq \Sigma^*$.
On the basis of membership and conjecture Queries, one tries to output $g= f$.

\paragraph{Angluin's Learning algorithm for regular sets}
The learning model is such that the teacher answers membership queries and
conjecture queries. Angluin's algorithm shows how to learn any regular set, i.e.
any function $\Sigma^*  \rightarrow \{0,1\}$, which is the characteristic
function of a regular set. It finds $f$ exactly, and the complexity of the procedure 
depends polynomially $O(m.n^2)$ on two parameters: $n$ the size of the minimum
automaton
for $f$ and $m$ the maximum length of counter examples returned by the conjecture
queries. Moreover there are at most $n$ conjecture Queries.

\paragraph{Learning without reset}
The Angluin model supposes a reset operator, similar to the reliable reset of section \ref{test},  but \cite{rs93}
showed  how to generalize the Angluin model without reset.
As seen in Section \ref{test}, a {\em homing sequence} is a sequence which uniquely
identifies the state after
reading the sequence. Every minimal deterministic finite automaton has a homing sequence $\sigma$.

The procedure runs  $n$ copies of Angluin's algorithm, $L_1,...,L_n$, where $L_i$
assumes
that $s_i$ is the initial state. After a membership query in $L_i$, one applies
the homing sequence $\sigma$, which leads to state $s_k$. One leaves $L_i$ and
continues in $L_k$.

 \subsection{Methods for  approximate decision problems}

In the previous section, we considered approximate methods for decision, counting and learning  problems. We now relax the decision and learning problems in order to 
obtain more efficient approximate methods.

 \subsubsection{Property testing}\label{pt}

Property testing is a statistics based approximation technique to decide if either
an input satisfies a given property, or is far from any input satisfying the
property, using only few samples of the input and a specific distance between inputs. It is later used in section
\ref{amc}.
The idea of moving the approximation to the
input was implicit in {\em Program Checking}~\cite{bk95,blr93,rs96},
in {\em Probabilistically Checkable Proofs} (PCP)~\cite{as98}, and explicitly studied for graph properties under the context of
property testing~\cite{ggr98}.
The class of sublinear algorithms has similar goals:
given a massive input, a sublinear algorithm can approximately decide a property by
sampling a tiny fraction of the input.
The design of sublinear algorithms is motivated by the recent considerable growth of
the size of the data that algorithms are called upon to process in everyday
real-time applications,
for example in bioinformatics for genome decoding or in Web 
databases for the search of documents.  
Linear-time, even polynomial-time, algorithms were considered 
to be efficient for a long time, but this is no longer the case, 
as inputs are vastly too large to be read in 
their entirety.  

Given a distance between objects, an $\eps$-tester for a property $P$ accepts all
inputs which satisfy the property and rejects with high probability all inputs which
are  $\eps$-far from inputs that satisfy the property. 
Inputs which are $\eps$-close to the property determine a gray area where no
guarantees exists. These restrictions allow for sublinear algorithms and even $O(1)$
time algorithms, whose complexity only depends on $\eps$. 

Let $\mathbf{K}$ be a class of finite structures with a normalized 
distance $\dist$ between structures, i.e.  $\dist$ lies
in $[0,1]$. 
For any $\eps>0$, we say that $U,U'\in\mathbf{K}$ are {\em $\eps$-close} if their
distance is at most $\eps$. They are {\em $\eps$-far} if they are not $\eps$-close.
In the classical setting, satisfiability is the decision problem 
whether $U\models P$ for a structure $U \in \mathbf{K}$  and a 
property $P\subset \mathbf{K}$.
A structure $U\in\mathbf{K}$ {\em $\eps$-satisfies} $P$, or  $U$ is $\eps$-close to $\mathbf{K}$
or $U \models_{\eps} P$ for short, if $U$ is $\eps$-close to some $U'\in\mathbf{K}$
such that $U'\models P$. We say that $U$ is $\eps$-far from $\mathbf{K}$
or $U \not\models_{\eps} P$ for short, if $U$ is not $\eps$-close to $\mathbf{K}$.
\begin{definition}[Property tester~\cite{ggr98}]
Let $\eps > 0$.
An {\em $\eps$-tester} for a property $P\subseteq\mathbf{K}$ is a randomized
algorithm $A$ such that, for any structure $U\in\mathbf{K}$ as input:\\
(1) If $U\models P$, then $A$ accepts;\\ 
(2) If $U\not\models_\eps P$, then $A$ rejects with probability at least $2/3$.%
\footnote{The constant $2/3$ can be replaced by any other constant $0<\gamma<1$ by 
iterating $O(\log (1/\gamma))$ the $\eps$-tester and accepting iff all the
executions accept}
\end{definition}

A {\em query} to an input structure $U$ depends on the model for accessing the
structure.
For a word $w$, a query asks for the value of $w[i]$, for some $i$.
For a tree $T$, a query asks for the value of the label of a node $i$,
and potentially for the label of its parent and its $j$-th successor, for some $j$. 
For a graph a query asks if there exists an edge between nodes $i$ and $j$.
We also assume that the algorithm may query the input size.
The {\em query complexity} is the number of  queries made to the structure.
The {\em time complexity} is the usual definition, where we assume that the
following operations are performed in constant time: arithmetic operations, a
uniform random choice of an integer
from any finite range not larger than the input size, and a query to the input.

\begin{definition}
A property $P\subseteq\mathbf{K}$ is {\em testable},  if there exists
 a randomized algorithm $A$  such that, for every real $\eps>0$ as input,
 $A(\eps)$ is an $\eps$-tester of $P$ whose 
 query and time complexities depend only on $\eps$ (and not on the input size).
\end{definition}

Tools based on property testing use an 
approximation on  inputs which allows to:
\begin{enumerate}
\item Reduce the decision of some global properties to the decision of local
properties by sampling,
\item Compress a structure to a constant size sketch on which a class of properties can
be approximated.
\end{enumerate}

We detail some of the methods on graphs, words and trees.
\paragraph{Graphs}
In the context of undirected graphs~\cite{ggr98}, the distance is the (normalized)
{\em Edit distance} on edges: 
the distance between two graphs on $n$ nodes
is the minimal number of edge-insertions and edge-deletions 
needed to modify one graph into the other one.
Let us consider the adjacency matrix model.
Therefore, a graph $G=(V,E)$
is said to be $\varepsilon$-close to another graph $G'$,
if $G$ is at distance at most $\varepsilon n^2$ from $G'$, that is
if $G$ differs from $G'$ in at most $\varepsilon n^2$ edges.

In several cases, the proof of testability of a graph property on the initial graph
is based on a reduction to a graph property on constant size but random subgraphs.
This was generalized for every testable graph properties by ~\cite{gt03}.
The notion of $\eps$-reducibility highlights this idea. 
For every graph $G=(V,E)$ and integer $k \geq 1$, let $\Pi$ denote the set
of all subsets $\pi\subseteq V$ of size $k$.
Denote by $G_{\pi}$ the vertex-induced subgraph of $G$ on $\pi$.
\begin{definition}\label{eps-reduc}
Let $\eps> 0$ be a real, $k\geq 1$ an integer,
and $\phi,\psi$ two graph properties.
Then $\phi$ is {\em $(\eps,k)$-reducible to $\psi$} if and only if
for every graph $G$,
\begin{eqnarray*}
\label{eq.reduc.1}
G\models\phi&\implies&\forall \pi\in\Pi,\ G_{\pi}\models\psi,\\
\label{eq.reduc.2}
G\not\models_{\eps}\phi&\implies&
\Pr_{\pi\in\Pi}  [ G_{\pi}\not\models\psi ]\geq 2/3.
\end{eqnarray*}
\end{definition}
Note that the second implication means that if $G$ is $\eps$-far to all graphs
satisfying the property $\phi$, then with probability at least $2/3$
a random subgraph on $k$ vertices does not satisfy $\psi$.

Therefore, in order to distinguish between a graph satisfying $\phi$
to another one that is far from all graphs satisfying $\phi$,
we only have to estimate the probability $\Pr_{\pi\in\Pi}  [ G_{\pi}\models\psi ]$.
In the first case, the probability is $1$, and in the second it is at most $1/3$.
This proves that the following generic test is an $\eps$-tester:\\
\begin{center}\begin{fmpage}{9cm}
\textbf{Generic Test$(\psi,\eps,k)$}\\
\begin{enumerate}
\item Input: A graph $G=(V,E)$
\item Generate uniformly a random subset $\pi\subseteq V$ of size $k$
\item Accept if $G_\pi\models \psi$ and reject otherwise
\end{enumerate}
\end{fmpage}
\end{center}
\begin{proposition}
If for every $\eps>0$, there exists $k_\eps$ such that $\phi$
is $(\eps,k_\eps)$-reducible to $\psi$, then the property
$\phi$ is testable. 
Moreover, for every $\eps>0$, \textbf{Generic Test$(\psi,\eps,k_\eps)$}
is an $\eps$-tester for $\phi$ whose query and time complexities are in $(k_\eps)^2$.
\end{proposition}

In fact, there is a converse of that result, and for instance
we can recast the testability of  
$c$-colorability~\cite{ggr98,ak02} in terms 
of $\eps$-reducibility.
Note that this result is quite surprising since $c$-colorability is an  NP-complete
problem for $c\geq 3$.
\begin{theorem}[\cite{ak02}]\label{ggr1}
For all $c\geq 2$, $\eps>0$,
$c$-colorability is $(\eps,O((c\ln c)/\eps^2))$-reducible
to $c$-colorability.
\end{theorem}

\paragraph{Words and trees}
\label{ptwt}
Property testing of regular languages was first considered in~\cite{akns00} for the
{\em Hamming distance},
and then extended to languages recognizable by bounded width read-once branching
programs~\cite{new02}, where the Hamming distance between two words is the
minimal number of character substitutions required
to transform one word into the other. 
The (normalized) edit distance between two words (resp. trees) of size $n$
is the minimal number of insertions, deletions
and substitutions of a letter (resp. node) required to transform one word (resp. tree)
into the other, divided by $n$.
When words are infinite, the distance is defined as the superior limit of the
distance of the
respective prefixes.

\cite{mr04} considered the testability of regular languages on words and trees
under the edit distance with {\em moves}, that 
considers one additional operation:
moving one arbitrary substring (resp. subtree) to another position in one step.
This distance seems to be more adapted in the context of property testing, since their
tester is more efficient and simpler  than the one of~\cite{akns00}, 
and can be generalized to tree regular languages.

\cite{fmr2010} developed a statistical embedding of words 
which has similarities with the Parikh mapping~\cite{par66}.
This embedding associate to every word a sketch of constant size (for fixed $\eps$)
which allows to decide any property given by some regular grammar or even some
context-free grammar.
This embedding has other implications that we will develop further
in Section~\ref{approxpt}.

\subsubsection{PAC and statistical learning}

The {\em Probably Approximately Correct} (PAC) learning model, introduced by Valiant \cite{val1} is a
framework to approximately learn an unknown function $f$ in a class $\cal{F}$, such that
each $f$ has a finite representation, i.e. a formula which defines $f$.
The model supposes positive
and negative samples along a distribution $\cal{D}$, i.e. values
$x_i,f(x_i)$ for $i=1,2,...,N$. The learning algorithm proposes a function
$h$ and the error between $f$ and $h$ along the distribution $\cal{D}$ is:
$$error(h)=\B_{x \in \cal{D}} [f(x) \neq h(x)]$$

A class  $\cal{F}$ of  functions $f$ is PAC-learnable if there is a
randomized  algorithm such
that for all $f \in  \cal{F},  \eps, \delta, \cal{D}$, it produces with
probability greater than $1-\delta$, an estimate $h$ for $f$ such that 
$error(h)\leq \eps$. It is efficiently PAC-learnable if the algorithm
is polynomial in $N, \frac{1}{\eps}, \frac{1}{\delta}, size(f)$, 
where $size(f)$ 
is the size of the finite representation of $f$. Such learning methods are
independent of the distribution $\cal{D}$, and are used in black box checking in
section \ref{bbc} to verify  a property of a black box by learning a model.

The class $\cal{H}$ of the functions $h$ is called the Hypothesis space and 
 the class is {\em properly learnable} if $\cal{H}$ is
 identical to $\cal{F}$:

\begin{itemize}
\item Regular languages are PAC-learnable. Just replace in Angluin's model, the 
conjecture queries by PAC queries, i.e. samples from  a distribution $\cal{D}$.
 Given a proposal $L'$ for $L$, we take N samples along $ \cal{D}$ and may obtain a 
counterexample, i.e. an element $x$ on which $L$ and $L'$ disagree. If $n$ is the
minimum number of states of the unknown $L$, then Angluin's algorithm with at most
$$N=O((n+1/\eps).(n \ln (1/\delta)+n^2)$$ samples can replace the $n$ conjecture
queries and 
guarantee with probability at least $1-\delta$ that the error is less than $\eps$.
\item $k$-DNF and $k$-CNF are learnable but it is not known whether CNF or DNF are
learnable.

\end{itemize}

The Vapnik-Chernovenkis (VC)  dimension \cite{Vapnik:81} of a class 
$\cal{F}$, denoted $VC( \cal{F})$
is the largest cardinality $d$ of a sample set $S$ that is shattered by $\cal{F}$,
i.e. such that for
every subset $S' \subseteq S$ there is  an $f \in \cal{F}$ such that $f(x)=a$ for
$x\in S'$,  $f(x)=b$ for $x\in S-S'$ and $a\neq b$.
 
A classical result
of \cite{behw,kv} is that if $d$ is finite then the class is PAC learnable.
If $N\geq O(\frac{1}{\eps}.\log \frac{1}{\delta}+\frac{d}{\eps}.\log \frac{1}{_eps}
)$, then any $h$ which is consistent with the samples, i.e. gives the same result as
$f$ on the random samples,  is a good estimate.
Statistical learning \cite{Vapnik:83} generalizes this approach from functions
to distributions.

\section{Applications to model checking and testing}

\subsection{Bounded and unbounded  model checking}\label{bmc}

 Recall that the {\em Model Checking} problem is to decide, given a transition system
$\mathcal{M}$ with an initial state $s_0$ and a temporal formula $\varphi$ whether
$\mathcal{M},s_0 \models \varphi$, i.e. if the system $\mathcal{M}$ satisfies the
property defined
by $\varphi$.
Bounded model checking introduced in \cite{bccz} is a useful method for detecting
errors, but incomplete in general for efficiency reasons: it may be intractable to
ensure that a property is satisfied. 
For example, if we consider some safety property expressed by a formula 
$\varphi=\mathbf{G} p$, $\mathcal{M},s_0 \models \mathbf{\forall}\varphi$ means that
all 
initialized paths in $\mathcal{M}$ satisfy $\varphi$, and $\mathcal{M},s_0 \models 
\mathbf{\exists}\varphi$ means that there exists an initialized path in 
$\mathcal{M}$ 
which  satisfies $\varphi$. Therefore, finding a counterexample to the property 
$\mathbf{G} p$ corresponds to the question whether there exists a path that is a 
witness for the property $\mathbf{F} \neg p$.

The basic idea of bounded model checking is to consider only a finite prefix of a
path that
may be a witness to an existential model checking problem. The length of
the 
prefix is restricted by some bound $k$. In practice, one progressively increases the bound, looking
for 
witnesses in longer and longer execution paths. A crucial observation is that,
though the
prefix of a path is finite, it represents an infinite path if there is a
\textit{back loop}
 to any of the previous states. If there is no such back loop, then the prefix does
not say
 anything about the infinite behavior of the path beyond state $s_k$.

The $k$-bounded semantics of model checking is defined by considering only finite
prefixes of
a path, with length $k$, and is an approximation to the unbounded semantics. We will
denote 
satisfaction with respect to the $k$-bounded semantics  by $\models_k$. The main
result of 
bounded model checking is the following.

\begin{theorem}
Let $\varphi$ be an LTL formula and  $\mathcal{M}$ be a transition system. Then
$\mathcal{M} 
\models \mathbf{\exists} \varphi$ iff there exists $k =
0(|\mathcal{M}|.2^{|\varphi|})$ such that 
$\mathcal{M} \models_k \mathbf{\exists} \varphi$.
\end{theorem}

Given a model checking problem $\mathcal{M} \models \mathbf{\exists} \varphi$, a
typical application of
 BMC starts at bound $0$ and increments the bound $k$ until a witness is found. This
represents a 
partial decision procedure for model checking problems:

\begin{itemize}
\item if $\mathcal{M} \models \mathbf{\exists} \varphi$, a witness of length $k$
exists, and the 
procedure terminates at length $k$.
\item if $\mathcal{M} \not\models \mathbf{\exists} \varphi$, the procedure does not
terminate.
\end{itemize}

For every finite transition system $\mathcal{M}$ and LTL formula $\phi$, there
exists a number $k$ 
such that the absence of errors up to $k$ proves that $\mathcal{M} \models
\mathbf{\forall}\phi$. 
We call $k$ the \textit{completeness treshold} of $\mathcal{M}$ with respect to $\phi$.
For example, the completeness treshold for a safety property expressed by a formula
$\mathbf{G} p$ 
is the minimal number of steps required to reach all states: it is the longest
``shortest path'' from 
an initial state to any reachable state.

In the case of unbounded model checking, one can formulate the check for property satisfaction as a SAT problem. A  general SAT approach \cite{abe2000} can be used for reachability analysis, when
the binary relation $R$ is represented by a {\em Reduced Boolean Circuit} (RBC), a specific 
logical circuit
with $\wedge, \neg, \leftrightarrow$. One can associate a SAT formula with the
binary  relation $R$ and each
$R^i$ which defines the states reachable at stage $i$ from $s_0$, i.e.
$R^0=\{s_0\}$, $R^{i+1}=\{s: \exists v R^i(v)\wedge vRs \}$. Reachability analysis
consists in computing unary sets $T^i$, for $i=1,...,m$: 
\begin{itemize}
\item $T^i$ is the set of states reachable at stage $i$ which satisfy
a predicate $Bad$, i.e. $\exists s (Bad(s)\wedge R^i( s))$,
\item compute $T^{i+1}$ and check if $T^i \leftrightarrow T^{i+1}$.
\end{itemize}

In some cases, one may have a  more succinct representation of the transitive closure 
of $R$.  A $SAT$ solver is used to perform all the decisions. 
 
\subsubsection{Translation  of BMC to SAT}

It remains to show how to reduce bounded model checking to propositional
satisfiability. 
This reduction enables to use efficient propositional SAT solvers to perform model
checking.
Given a transition system $\mathcal{M}= (S,I,R,L)$ where $I$ is the set of initial states, an LTL formula $\varphi$ and a
bound $k$, one can 
construct a propositional formula $[\mathcal{M},\varphi ]_k$ such that:

$$\mathcal{M} \models_k \mathbf{\exists} \varphi \mbox{ iff } [\mathcal{M},\varphi
]_k \mbox{ is satisfiable}$$

Let $(s_0,\dots,s_k)$ the finite prefix, of length $k$, of a path $\sigma$. Each $s_i$ 
represents a state at time step $i$ and consists of an assignment of truth values to
the set 
of state variables. The formula $[\mathcal{M},\varphi ]_k$ encodes constraints on 
$(s_0,\dots,s_k)$ such that $[\mathcal{M},\varphi ]_k$ is satisfiable iff $\sigma$
is a 
witness for $\varphi$. 

The first part $[\mathcal{M}]_k$ of the translation is a propositional formula that
forces  
$(s_0,\dots,s_k)$ to be a path starting from an initial state: 
$[\mathcal{M}]_k = I(s_0) \wedge \bigwedge_{i=0}^{k-1} R(s_i,s_{i+1})$.

The second part $[\varphi ]_k$ is a propositional formula which means that $\sigma$
satisfies $\varphi$
 for the $k$-bounded semantics. For example, if $\varphi$ is the formula $\mathbf{F}
 p$, the formula 
 $[\varphi ]_k$ is simply the formula $\bigvee_{i=0}^k p(s_i)$. In general,
the second part of the translation depends on the shape of the path $\sigma$:
 
\begin{itemize}
\item If $\sigma$ is a $k$-loop, i.e. if there is a transition from state $s_k$ to a
state 
$s_l$ with $l\leq k$, we can define a formula $[\varphi ]_{k,l}$, by induction on
$\varphi$,
 such that the formula $ \bigvee_{l=0}^k (R(s_k,s_l) \wedge [\varphi ]_{k,l})$ means
that 
$\sigma$ satisfies $\varphi$.
\item If $\sigma$ is not a $k$-loop, we can define a formula $[\varphi ]_k$, by
induction on 
$\varphi$, such that the formula $ (\neg \bigvee_{l=0}^k R(s_k,s_l)) \wedge
[\varphi ]_k$
means that $\sigma$ satisfies $\varphi$ for the $k$-bounded semantics.
\end{itemize}
We now explain how interpolation can be used to improve the efficiency of  SAT based bounded model
checking.

\subsubsection{Interpolation in propositional logic}
{\em Craig's interpolation } theorem is a fundamental result of mathematical logic. For propositional 
formulas $A$ and $B$, if 
$A \rightarrow B$, there is a formula $A'$ in the common language of $A,B$
such that $A \rightarrow A'$ and  $A' \rightarrow B$. 
Example: $A=p \wedge q$, $B=q \vee r$. Then $A'=q$.

In the model checking context, \cite{mcm03} proposed to  use the interpolation as
follows. Consider formulas $A,B$ in CNF normal form, and let $(A,B)$ be the set of clauses of $A$ and $B$. 
Instead of showing $A \rightarrow C$, we set $B=\neg C$ and show that $(A,B)$ is unsatisfiable.

If $(A,B)$ is unsatisfiable, we apply Craig's theorem and conclude that there is an $A'$ such that
$A \rightarrow A'$ and  $(A', B)$  is unsatisfiable.
Suppose $A$ is the set of clauses associated with an  automaton or a transition
system and $B$ is the set of clauses associated with the negation of the formula to
be checked. 
Then $A'$ defines the possible errors. 

\begin{figure}[h]
\centerline{\includegraphics[width=10cm]{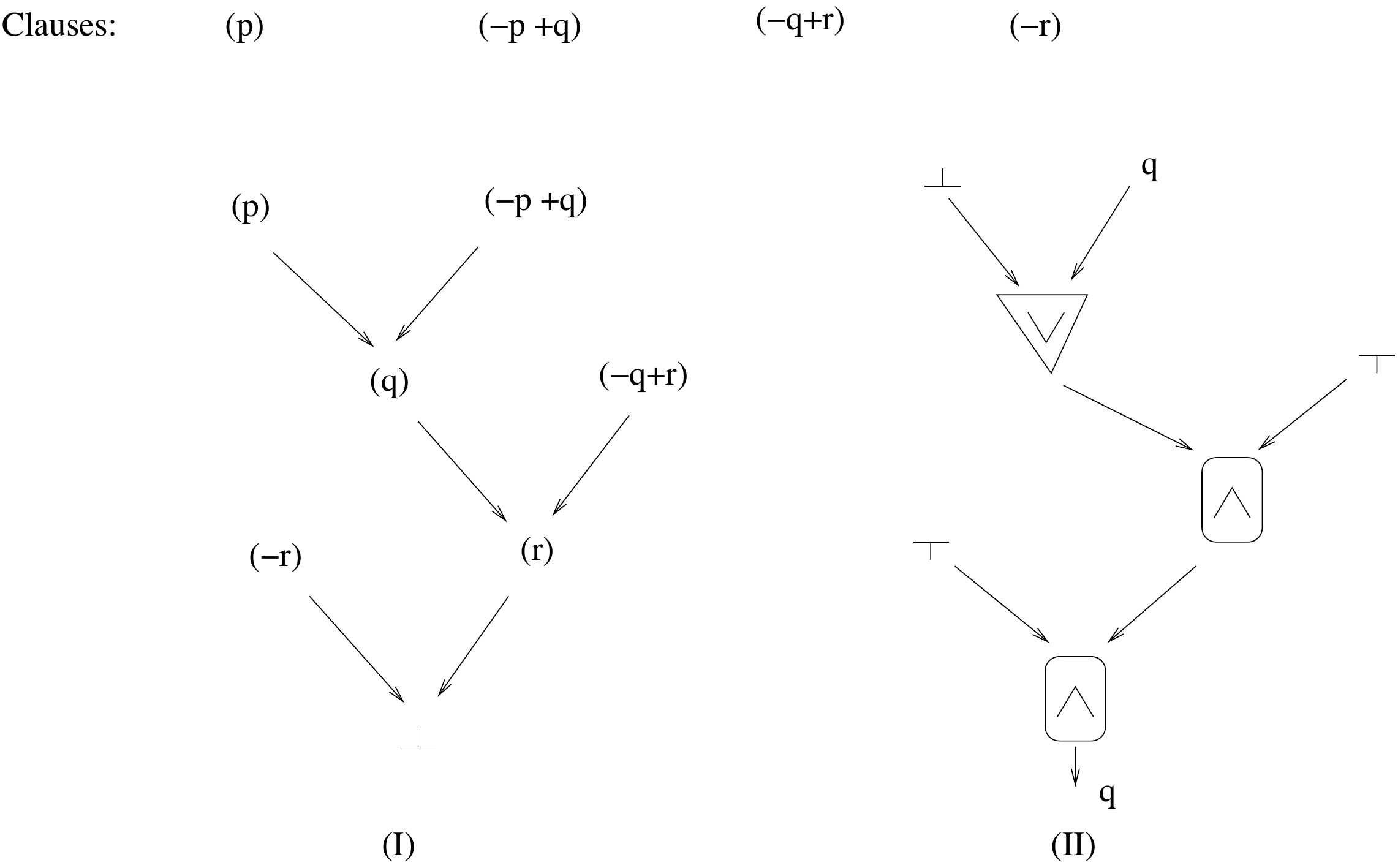}}
\caption{Craig Interpolant: $A: \{(p), (\neg p \vee q)\}$, and $B:\{(\neg q \vee r
),(\neg r)\}$. The proof by resolution (I) shows that $(A,B)$ is unsatisfiable. The
circuit (II) (with OR and AND gates, input labels which depend on the clauses, as explained in definition \ref{int})
mimics the proof by resolution and output the interpolant $A'=q$. }\label{craig}
\end{figure}

There is a direct connection between a resolution proof of the unsatisfiability of $(A,B)$
and the interpolant $A'$. It suffices to keep the same structure of the resolution proof and only modify the labels, as
 explained in Figure \ref{craig}.

{\bf Resolution rule}.
Given two clauses  $\mathcal{C}_1,\mathcal{C}_2$ such that a variable $p$ appears positively in $\mathcal{C}_1$ and negatively in $\mathcal{C}_2$, i.e. $\mathcal{C}_1=p\vee \mathcal{C}'_1$ and $\mathcal{C}_2=\neg p\vee \mathcal{C}'_2$, the {\em resolution rule} on the {\em pivot} $p$ yields the 
{\em resolvent}  $\mathcal{C'}=\mathcal{C}'_1 \vee \mathcal{C}'_2$.
If the two clauses are  $\mathcal{C}_1=p$ and $\mathcal{C}_2=\neg p$, the resolvent on pivot $p$ is $\bot$ (the symbol for false).
The  proof $\Pi$ of unsatisfiability of $(A,B)$ by resolution, can be
represented by a {\em Directed Acyclic Graph} DAG with labels on the nodes, several input nodes and one output node, as in figure  \ref{craig}-(I).
Clauses of $A,B$ are labels of the input nodes, clauses $\mathcal{C'}$ obtained by one application of the resolution rule 
are labels of the internal nodes,
and  $\bot$ is the label of the unique output nodes.

{\bf Obtaining an interpolant}.
For the sets of clauses $(A,B)$, a variable $v$ is global if it
occurs both in $A$ and in $B$, otherwise it is local to $A$ or to $B$. 
Let $g$ be a function which transforms a clause into another clause.
For a clause $\mathcal{C}\in
A$, let
$g(\mathcal{C})$ be the disjunction of its global litterals, let 
$g(\mathcal{C})=\bot$ (false) if no global litteral is present and let 
$g(\mathcal{C})=\top$ (true) if $\mathcal{C}\in
B$. 

The labels of the internal nodes and output node are specified by  definition \ref{int}, on a copy $\Pi'$ of $\Pi$. 

\begin{definition}\label{int}
For all labels $\mathcal{C}$ of nodes of $\Pi$, let $\mu(\mathcal{C})$ be a boolean formula which is the new label of
$\Pi'$.
\begin{itemize}\label{int}
\item if $\mathcal{C}$ is the label of an input node then 
$\mu(\mathcal{C})=g(\mathcal{C})$.
\item let $\mathcal{C}$ be a resolvent on $\mathcal{C}_1,\mathcal{C}_2$ using the
pivot $p$:  if $p$
is local to $A$, then $\mu(\mathcal{C})=p(\mathcal{C}_1)\vee p(\mathcal{C}_2)$ otherwise
$\mu(\mathcal{C})=p(\mathcal{C}_1)\wedge p(\mathcal{C}_2)$
\end{itemize}
The interpolant of  $(A,B)$ along $\Pi$ is $\mu(\bot)$, i.e. the clause associated with the
DAG's unique output node.
\end{definition}
This construction yields a direct method to obtain an interpolant from an
an unsatisfiability proof. It isolates a subset of the clauses from $A,B$, which can
be viewed as an {\em abstraction of the unsatisfiability proof}.
This approach 
 is
developped further in \cite{JMM04}.

\subsubsection{Interpolation and SAT based model checking}

One can formulate  the problem of safety property verification in the following terms 
\cite{mcm03}.
Let $\mathcal{M} = (S,R,I,L)$ be a transition system  and $F$  a final constraint.
The initial constraint $I$, the final constraint 
$F$ and the transition relation $R$ are expressed by propositional formulas over
boolean variables (a state is represented by a truth assignment for $n$ variables
$(v_1,\dots,v_n)$). 


An accepting path of  $\mathcal{M}$ 
is a sequence of states  $(s_0,\dots,s_k)$ such that the formula $I(s_0) \wedge
(\bigwedge_{i=0}^{k-1} R(s_i,s_{i+1})) \wedge F(s_k)$ is true. 
In bounded model checking, one translates the existence of an accepting path 
of length $0\leq i \leq k+1$ into a propositional satisfiability problem by
introducing a new indexed set
of variables $W_i= \{w_{i1},\dots,w_{in}\}$ for $0\leq i \leq k+1$. 
An accepting path of length in the range 
$\{0,\dots,k+1\}$ exists exactly when the following formula is satisfiable:
$$bmc_0^k = I(W_0) \wedge (\bigwedge_{i=0}^{k} R(W_i,W_{i+1})) \wedge
(\bigvee_{i=0}^{k+1} F(W_i))$$
In order to apply the interpolation technique, one expresses the existence of a prefix of length
$1$ and of a suffix of length $k$ by the following formulas:
$$\textit{pre}_1(\mathcal{M})=  I(W_0) \wedge R(W_0,W_1)$$
$$\textit{suf}_1^k(\mathcal{M}) = (\bigwedge_{i=1}^{k} R(W_i,W_{i+1})) \wedge
(\bigvee_{i=1}^{k+1} F(W_i))$$
 
To apply a SAT solver, one assumes the existence of some function CNF that translates
a boolean formula $f$ into a set of clauses $CNF(f,U)$ where $U$ is a set of fresh
variables, not occurring in $f$. Given two sets of clauses $A,B$ such that $A\cup B$
is unsatisfiable and a proof $\Pi$ of unsatisfiability, we note $Interpolant (\Pi,
A,B)$ the associated interpolant.
Below, we give a procedure to check the existence of a finite accepting path of
$\mathcal{M}$, introduced in \cite{mcm03}. 
The procedure is parametrized by a fixed value $k \geq 0$.\\

\begin{algo}

Procedure FiniteRun($M=(I,R,F),k$)\\

 if $(I\wedge F)$ is satisfiable, return \textit{True}

 let $T=I$

 while (true)

 let $M'=(T,R,F)$,  $A=CNF(\textit{pre}_1(M'),U_1)$, $B=CNF(\textit{suf}_1^k(M'),U_2)$

 Run SAT on $A\cup B$

 If ($A\cup B$ is satisfiable) then

 ~~~if $T=I$ then return True else abort

else (if $A\cup B$ unsatisfiable)

 ~~~let $\Pi$ be a proof of unsatisfiability of $A\cup B$, $P= Interpolant (\Pi,
A,B)$, $T'= P(W/W_O)$

 ~~~if $T'$ implies $T$ return \textit{False}

 ~~~let $T= T\cup T'$

endwhile

end
\end{algo}


\begin{theorem}(\cite{mcm03})
For $k>0$, if FiniteRun($\mathcal{M},k$) terminates without aborting, it returns
\textit{True} iff
$\mathcal{M}$ has an accepting path.
\end{theorem}

This procedure terminates for sufficiently large values of $k$: the \textit{reverse
depth} of $\mathcal{M}$ is the maximum length of the shortest path from any state to
a 
state satisfying $F$. When the procedure aborts, one only has to increase the value
of $k$. Eventually the procedure will terminate. Using
interpolation in SAT based model checking is a way to complete and accelerate
bounded model checking.

\subsection{Approximate model checking}\label{amc}

We first consider a heuristics (Monte-Carlo) to verify an LTL formula, and then consider two methods where both approximation and randomness are used to obtain probabilistic abstractions, based on property and equivalence testers.

\subsubsection{Monte-Carlo model checking}

\label{principle} 

In this section, we present a randomized Monte-Carlo algorithm for linear temporal
logic 
model checking \cite{grosu}.
Given a deterministic transition system $\mathcal{M}$ and a temporal logic formula
$\phi$,
the model checking problem is to decide whether $\mathcal{M}$ satisfies $\phi$. In
case  
$\phi$ is linear temporal logic (LTL) formula, the problem can be solved by reducing
it to 
the language emptiness problem for finite automata over infinite words \cite{vw}. The 
reduction involves modeling $\mathcal{M}$ and $\neg\phi$ as B\"uchi automata 
$A_{\mathcal{M}}$ and $A_{\neg\phi}$, taking the product  $A=A_{\mathcal{M}} \times 
A_{\neg\phi}$, and checking whether the language $L(A)$ of $A$ is empty. Each  LTL
formula 
$\phi$ can be translated to a B\"uchi automaton whose language is the set of
infinite words 
satisfying $\phi$ by using a tableau construction.\\
The presence in $A$ of an accepting lasso, where a {\em lasso} is  a cycle reachable
from an 
initial state of $A$, means that $\mathcal{M}$ is not a model of $\phi$.
  
\paragraph{Estimation method.}
\label{estimation}

To each instance $\mathcal{M} \models \phi$ of the LTL model checking problem, one may 
associate a Bernouilli random variable $z$ that takes  value $1$ with probability
$p_Z$ 
and value $0$ with probability $1-p_Z$. Intuitively, $p_Z$ is the probability that an 
arbitrary execution path of $\mathcal{M}$ is a counterexample to $\phi$. Since $p_Z$
is 
hard to compute, one can use a Monte-Carlo method to derive a one-sided error
randomized 
algorithm for LTL model checking.

Given a Bernouilli random variable $Z$, define the geometric  random variable $X$ with 
parameter $p_Z$ whose value is the number of independent trials required until success.
The probability distribution of $X$ is:
$$p(N)=Pr[X=N]=q_Z^{N-1}.p_Z$$ where $q_z = 1- p_z$,
 and the cumulative distribution is 
$$Pr[X \leq N]=\sum_{n=0}^N p(n) = 1-q_Z^N$$
Requiring that $Pr[X \leq N]= 1- \delta$ for confidence ratio $\delta$ yields:
$N = ln(\delta)/ln(1-p_Z)$ which provides the number of attempts $N$ needed to achieve 
success with probability greater $1- \delta$.  Given an error margin 
$\varepsilon$ and assuming the hypothesis $p_Z \geq \varepsilon$, we obtain that:\\
$M = ln(\delta)/ln(1-\varepsilon) \geq ln(\delta)/ln(1-p_Z)$ and 
$Pr[X \leq M] \geq Pr[X \leq N] \geq 1- \delta$.\\
Thus $M$ is the minimal number of attempts needed to achieve success with confidence
ratio
$\delta$, under the assumption $p_Z \geq \varepsilon$. 

\paragraph{Monte-Carlo algorithm.}
\label{MC2}

The $\textrm{MC}^2$ algorithm samples lassos in the automaton $A$ via a random walk
through 
$A$'s transition graph, starting from a randomly selected initial state of $A$, and
decides
 if the cycle contains an accepting state.
 

\begin{definition} 
A finite run $\sigma = s_0 x_0 s_1 x_1\dots s_n x_n s_{n+1}$ of a B\"uchi automaton 
$A= (\Sigma, S, s_0, R, F)$ is called a lasso if $s_0,\dots,s_n$ are pairwise
distinct 
and $s_{n+1}=s_i$  for some $0\leq i \leq n.$ Moreover, $\sigma$ is said an
accepting lasso
 if some $s_j \in F$ ($i \leq j \leq n$), otherwise it is a non accepting lasso.
The lasso sample space $L$ of $A$ is the set of all lassos of $A$, while $L_a$ and
$L_n$ 
 are the sets of all accepting and non accepting lassos of $A$, respectively.
\end{definition}

To obtain a probability space over $L$, we define the probability of a lasso.

\begin{definition}
The probability $Pr[\sigma]$ of a finite run $\sigma = s_0x_0\dots s_{n-1}x_ns_n$ 
of a 
B\"uchi automaton $A$ is defined inductively as follows:
$Pr[s_0]= 1/k$ if $|s_0| = k$ and $Pr[s_0x_0x_1\dots s_{n-1}x_ns_n] = 
Pr[s_0x_0\dots s_{n-1}].\pi(s_{n-1},x_n, s_n)$ where $\pi(s,x,t)= 1/m$ if $(s,x,t) \in
R$ and
$|R (s)|=m$. Recall that $R(s)=\{ t : \exists x \in \Sigma, (s,x,t) \in R \}$. 
\end{definition}
Note that the above definition explores uniformly outgoing transitions
and corresponds to a random walk on the probabilistic space of lassos.

\begin{proposition}
Given a B\"uchi automaton $A$, the pair $(\mathcal{P}(L), Pr)$ defines a discrete
probability 
space.
\end{proposition}

\begin{definition}
The random variable $Z$ associated with the probability space $\mathcal{P}(L), Pr$
is defined
 by: $p_Z = Pr[Z=1] = \sum_{\sigma \in L_a} Pr[\sigma]$ and $q_Z = Pr[Z=0] =
\sum_{\sigma 
\in L_n} Pr[\sigma]$.
\end{definition}

\begin{theorem}
Given a B\"uchi automaton $A$ and parameters $\varepsilon$ and $\delta$ if
$\textrm{MC}^2$
returns false, then $L(A) \neq \emptyset $. Otherwise, $Pr[X>M|H_0]< \delta$ where 
$M = ln(\delta)/ln(1-\varepsilon)$ and $H_0 \equiv p_Z \geq \varepsilon$.
\end{theorem}

This Monte-Carlo decision procedure has time complexity O(M.D) and
space complexity O(D), where D is the diameter of the B\"uchi product
automaton.

This approach by statistical hypothesis testing for classical LTL model checking has
an 
important drawback: if $0<p_Z<\varepsilon$, there is no guarantee to find a
corresponding
 error trace. However, it would be possible to improve the quality of the result of the
 random walk by randomly reinitializing the origin of each random path in the
connected 
component of the initial state.

\subsubsection{Probabilistic abstraction} 

Symbolic model checking~\cite{macm,cgp} uses a succinct representation of a
transition system, such as an  ordered binary decision diagrams (OBDD)
~\cite{br86,b91} or a SAT instance.
In some cases, such as programs for integer multiplication or bipartiteness, the
OBDD size remains exponential.
The abstraction method (see Section \ref{aia})
provides a solution in some cases, when the OBDD size is intractable. We now
consider random substructures $(\widehat{\mathcal{M}})_\pi$ of finite size,
where $\pi$ denotes the random parameter, and study cases
when  we can infer a specification SPEC in an approximate way, by
checking whether random abstractions $\pi$ satisfy 
with sufficiently good probability (say $1/2$) 
on the choice of $\pi$, another specification SPEC' which depends on SPEC  and $\pi$.

We have seen in section \ref{pt} on property testing, that many graph properties on
large graphs are $\eps$-reducible to other graph properties on a random subgraph of
constant size.
Recall that  a graph property $\phi$ is {\em $\eps$-reducible to $\psi $} if testing
$\psi$ on random subgraphs of constant size suffices to distinguish between graphs
which satisfy $\phi$, and those that are $\eps$-far from satisfying $ \phi$.
Based on those results, one can define the concept of probabilistic abstraction for
transition systems of deterministic programs whose purpose is to decide some graph
property.
Following this approach, \cite{llmpr} extended the range of abstractions to programs
for a large family  of graphs properties using randomized methods.
A {\em probabilistic abstraction} associates  small random transition systems, to a
program and to a property.
One can then  distinguish with sufficient confidence between programs that accept
only graphs that satisfy $\phi$ and those which accept some graph that is $\eps$-far
from any graph that satisfies $\phi$.

In particular, the abstraction method 
has been applied  to a program for  graph bipartiteness.
On the one hand, a probabilistic abstraction on a specific program for testing
bipartiteness and other temporal properties has been constructed such that the
related transition systems have constant size.
On the other hand, an abstraction was shown to be necessary, in the sense that the
relaxation of the test alone does not yield OBDDs small enough to use the standard
model checking method.
To illustrate the method, consider the following specification, where
$\phi$ is a graph property,
\begin{quote}
{\em SPEC: The program $P$ accepts only if the graph $G$ satisfies $\phi$.}
\end{quote}
The graph $G$ is described by some input variables of $P$
providing the values of the   adjacency matrix of $G$.
We consider a transition system $\mathcal{M}$ which represents $P$, parametrized 
by the graph input $G$.
The method remains valid for the more general specifications,
where $\Theta$ is in $\boldsymbol{\exists}\mathrm{CTL}^{*}$,
\begin{quote}
{\em SPEC: $\mathcal{M},G \models \Theta$ only if $G$ satisfies $\phi$.}
\end{quote}

The formula $\Theta$, written in temporal logic, states that the program reaches an accepting state, on input $G$.
The states of $\mathcal{M}$ are determined by the variables and the constants of $P$.
The probabilistic abstraction is based on property testing.
Fix $k$ an integer, $\eps>0$ a real, and  another graph property $\psi$ such
that $\phi$ is $(\eps,k)$-reducible to $\psi$.
Let $\Pi$ be the collection of all vertex subsets of size $k$.
The probabilistic abstraction is defined for any random choice of $\pi\in \Pi$.
For all vertex subsets  $\pi\in\Pi$, consider any abstraction
$\widehat{\mathcal{M}}^{\pi}$ for the transition system $\mathcal{M}$  
such that the graph $G$ is abstracted to its restriction on $\pi$,
that we denote by $G_{\pi}$.
The abstraction of the formula $\Theta$ is done according to the transformation
$\mathcal{D}$, defined in Section~\ref{aia}.

We now present the generic probabilistic tester based on the above abstraction.
\begin{center}
\begin{fmpage}{9cm}
\textbf{Graph Test$\big(
(\Pi,\mathcal{M}),\Theta,\psi\big)$}\\
\begin{enumerate}
\item Randomly choose a vertex subset $\pi\in\Pi$.
\item Accept iff
\quad{$\forall {G}_{\pi} \quad 
(\widehat{\mathcal{M}}^{\pi} \models
\mathcal{D}(\Theta)\quad\implies\quad
{G}_{\pi}\models\psi).$}
\end{enumerate}
\end{fmpage}
\end{center}
The following theorem states the validity of the abstraction.
\begin{theorem}\label{generic-rob}
Let  $\Theta$ be in $\boldsymbol{\exists}\mathrm{CTL}^{*}$.
Let $\eps>0$ be a real, $k \geq 1$ an integer, and
$\phi$ be a formula $(\eps,k)$-reducible to $\psi$.
If there exists a graph $G$ such that
$\mathcal{M},G\models \Theta$ and $G\not\models_{\eps} \phi$,
then \textbf{Graph Test$\big(
(\Pi,\mathcal{M}),\Theta,\psi\big)$} rejects
with probability at least $2/3$.
\end{theorem}

This approximate method has a time complexity independent of $n$, the size of the structure, and only dependent on $\eps$.

\subsubsection{Approximate abstraction}\label{approxpt}

In \cite{fmr2010}, an equivalence tester is introduced and decides if two properties
are identical or $\eps$-far, i.e. if there is a structure which satisfies one
property but which is $\eps$-far from the other property, in time which only depends
on $\eps$. It generalizes property testing to {\em Equivalence Testing} in the case we want
 to distinguish two properties, and has direct applications for approximate model checking.

Two automata defining respectively two languages $L_1$ and $L_2$ are {\em
$\eps$-equivalent} when all but finitely many words $w\in L_1$ are $\eps$-close to 
$L_2$, and conversely.
The tester transform both  transition systems and a specification
(formula) into B\"uchi automata,
and test their approximate equivalence efficiently.
In fact, the $\eps$-equivalence of nondeterministic finite automata can be decided
in deterministic polynomial time, that is $m^{\size{\Sigma}^{O(1/ \eps)}}$ whereas
the exact decision version of this problem is PSPACE-complete by~\cite{ms73}, and in
deterministic exponential time algorithm for the $\eps$-equivalence testing of
context-free grammars, whereas the exact decision version is not  recursively
computable.

The comparison of two B\"uchi automata is realized by computing a constant size
sketch for each of them. The comparison is done directly on the sketches.
Therefore sketches are abstractions of the initial transition systems where
equivalence and implication can be approximately decided.
 More precisely, the sketch is an $\ell_1$-embedding of the language.
 Fix a B\"uchi automaton $A$.
 Consider all the (finite) loops of $A$ that contains an accepting state, and
compute the statistics
 of their subwords of length $1/\eps$. 
 The embedding $\mathcal{H}(A)$ is simply the set of these statistics. 
 The main result states that approximate equivalence on B\"uchi automata is
 characterized by the $\ell_1$-embedding in terms of statistics of their loops.
 \begin{theorem}
 Let $A,B$ be two B\"uchi automata.
 If $A$ and $B$ recognize the same language then
 $\mathcal{H}(A)=\mathcal{H}(B)$.
 If $A$ (respectively $B$) recognizes an infinite word $w$ such that
 $B$ (respectively $A$) does not recognize any word $\eps/4$-close to $w$, then
 $\mathcal{H}(A)\neq\mathcal{H}(B)$.
 \end{theorem}

This approximate method has a time complexity polynomial 
in the size of the automata.

\subsection{Approximate black box checking}\label{bbc}
Given a black box $A$, a {\em Conformance test} compares the black box to a model
$B$ for for a given conformance relation (cf Section \ref{nondet}), whereas {\em
Black Box Checking} verifies if 
the black box $A$ 
satisfies a property defined by a formula $\psi$.
When the conformance relation is the equivalence, conformance testing can use the 
Vasilevskii-Chow method \cite{Vasilevskii1973}, which remains an exponential
method $O(l^2.n.p^{n-l+1})$, where $l$ is the known number of states of the
model $B$, and $n$ is a known upper-bound for $|A |$ ($n \geq l$) and $p$ is the size of the
alphabet. 

\subsubsection{Heuristics for black box checking} 
\cite{PVY} proposes the following $O(p^n)$ strategy to check if  a
black box $A$ satisfies a property $\psi$.
They build a sequence of automata $M_1, M_2,...,M_i,...$ which converges to a model $B$
of $A$, refining Angluin's learning algorithm. 
The automaton $M_i$ is considered as a classical automaton and as a B\"uchi
automaton which accepts infinite words. 
Let $P$ be a B\"uchi automaton, introduced in section \ref{ba}, 
 associated with $\neg \psi$.
Given two B\"uchi automata, $P$ and $M_i$, one can use model checking to test if 
the intersection is empty, i.e. if $L(M_i) \cap L(P) = \emptyset$: this operation is exponential in the size of the automata.

If $L(M_i) \cap L(P) \neq \emptyset$, there is $\sigma_1,\sigma_2$ such that
$\sigma_1 . \sigma_2^{\infty}$ is in $M_i$ as a B\"uchi automaton and in $P$, and
such that
$\sigma_1 . \sigma_2^{n+1}$ is accepted by the classical $M_i$.
Apply $\sigma_1 . \sigma_2^{n+1}$ to $A$. If $A$ accepts there is  an error as 
$A$ also accepts  $\sigma_1 . \sigma_2^{\infty}$, i.e. an input which does not
satisfy the property. If $A$ rejects then
$M_i$ and $A$ differ and one can use Angluin's algorithm to learn $M_{i+1}$
from $M_i$ and the separating sequence $\sigma=\sigma_1 . \sigma_2^{n+1}$.

\begin{figure}[h]
\centerline{\includegraphics[width=16cm]{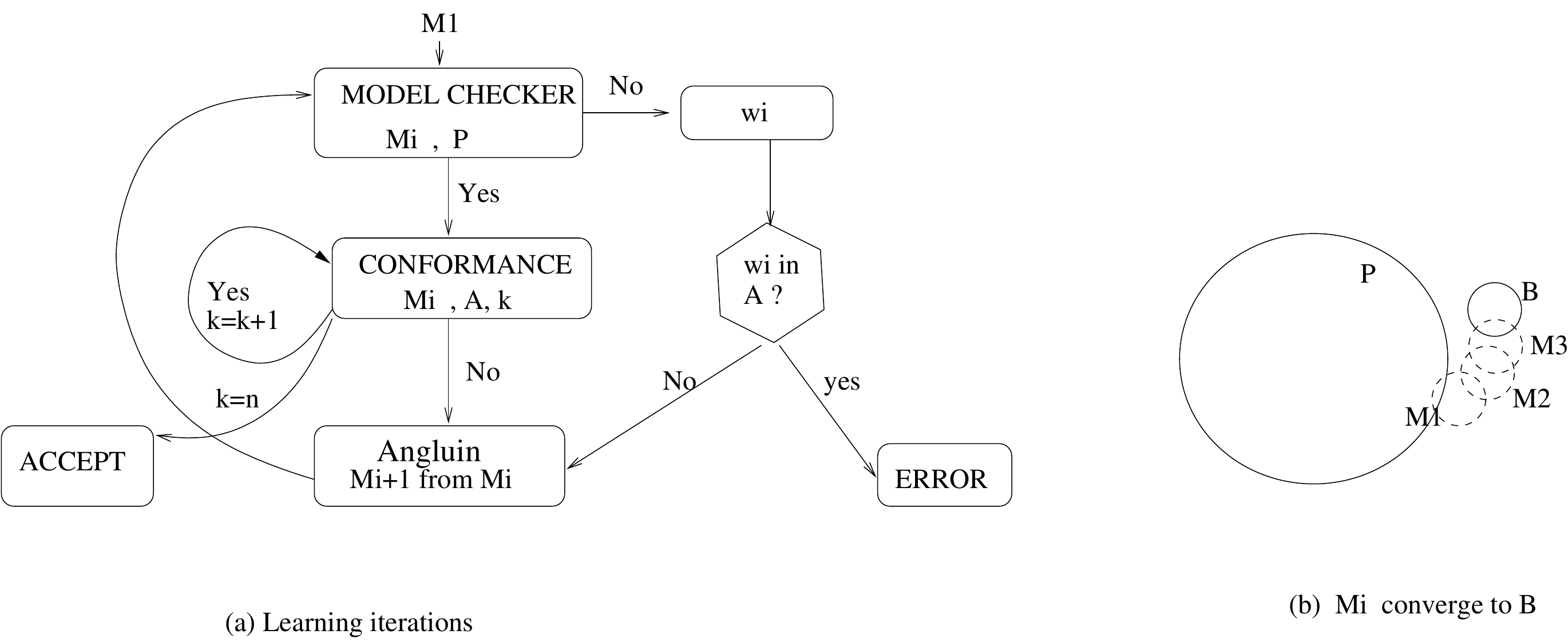}}
\caption{Peled-Vardi-Yanakakis learning Scheme in (a), and the sequence of $M_i$ in
(b).}\label{ys}
\end{figure}

If $L(M_i) \cap L(P) = \emptyset$, one can compare $M_i$ with $A$ using
Vasilevskii-Chow's conformance algorithm. If they
are different, the algorithm provides a sequence $\sigma$ where they differ
and one can use the learning
algorithm to propose $M_{i+1}$ with more states. If the conformance test succeeds and 
$k=| M_{i}|$, one keeps applying it with larger values of $k$, i.e. $k+1,...,n$.
 See Figure \ref{ys}. The pseudo-code of the procedure is:\\

{\bf Black box checking strategy $(A,P,n)$.}
\begin{itemize}
\item Set $L(M_1)=\emptyset$.


\item {\em Loop}: $L(M_i) \cap L(P) \neq \emptyset ~~?$ (model checking). 

\begin{itemize}
\item   If $L(M_i) \cap L(P) \neq \emptyset$, the intersection contains some
$\sigma_1 . \sigma_2^{\infty}$ such that $\sigma_1 . \sigma_2^{j} \in L(M_i)$ for
all finite $j$. Enter $w_i=reset. \sigma_1 . \sigma_2^{n+1}$ to $A$. If $A$ accepts
then there is an error as there is a word in $L(P) \cap L(A)$, then {\em Reject}. If
$A$ rejects then $A \neq M_i$, then go to {\em Learn $M_{i+1}(w_i)$}.

\item  If $L(M_i) \cap L(P) = \emptyset $.

{\em Conformance}: check whether $M_i$ of size $k$ 
conforms with $A$  with the Vasilevskii-Chow
algorithm with input $A,M_i,k$. If not, Vasilevskii-Chow provides a separating
sequence $\sigma$, then go to {\em Learn $M_{i+1}(\sigma)$}.
If $k=n$ then {\em Accept}, else set $k= k+1$ and go to {\em Conformance}.

\item {\em Learn $M_{i+1}(\sigma)$}: Apply Angluin algorithm from $M_i$ and the sequence 
$\sigma$ not in $M_i$.
Go to {\em Loop}.

\end{itemize}

\end{itemize}

This procedure uses model checking, conformance testing and learning. If one knows 
$B$, one
could  directly use the Vasilevskii-Chow algorithm
with input $A,B, n$ but it is exponential, i.e. $O(p^{n - l+1})$. With this strategy,
one tries to discover errors by approximating $A$ with $M_i$ with $k$ states and
hopes to catch errors earlier on.
The model checking
 step is exponential and the conformance testing is only exponential when $k >l$.
 

We could relax the black box checking, and consider close inputs, i.e. decide if an
input $x$
 accepted by $A$ is $\eps$ close to $\psi$ and hope for a polynomial algorithm in $n$.

\subsubsection{Approximate black box checking for close inputs} 

In  the previous Figure \ref{ys}, we can  relax the model checking step, exponential in $n$,
 by the approximate model checking, polynomial in $n$, as in section \ref{amc}.
Similarly, 
the conformance equivalence could be replaced by an approximate version where we consider close
inputs, i.e.
inputs with an edit distance with moves less than $\eps$. In this setting,
{\em Approximate Conformance} checks whether $M_i$ of size $k$ 
conforms within $\eps$ with $A$. 
It is an open problem
whether there exists a randomized algorithm, polynomial time in $n$, for {\em Approximate Conformance Testing}.

\subsection{Approximate model-based testing}\label{ambt}

In this subsection we first briefly present a class of methods that are, in some
sense, dual to the previous ones: 
observations from tests are used to learn partial models of components under tests,
from which further tests can be derived.
Then we present an  approach to random testing that is based on  uniform generation
and counting seen in Section \ref{counting}. 
It makes possible to define a notion of approximation of test coverage and to assess
the results
of a random test suite for such approximations.    

\subsubsection{Testing as learning partial models}

Similarities between testing and symbolic learning methods have been noticed since
the early eighties
\cite{BuddAngluin1982,CherniavskySmith1987}. Recently, this close relationship has
been formalized by Berg et al. 
in \cite{BergJonnson2005}.
However, the few reported attempts of using Angluin's-like inference algorithms for
testing have been faced to the difficulty
of implementing an oracle for the conjecture queries. 
Besides, Angluin's algorithm and its variants are limited to the learning of regular
sets: the underlying models are finite automata that are not well suited for
modeling software.

\cite{Groz2007} propose a testing method where model inference is used for black box
software components, combining unit testing (i.e. independent testing of each
component) and integration testing (i. e. global testing of the combined
components).
The inferred models are PFSM (Parametrized FSM), that are the following restriction
of  EFSMs (cf. Section \ref{efsm}): inputs and outputs can be parametrized by
variables, but not the states; transitions are labelled by some parametrized input,
some guard on these parameters, and some function that computes the output
corresponding to the input parameters. 

The method  alternates phases of model inference for each components, that follow
rather closely the construction of a conjecture in Angluin's algorithms, and phases
of model-based testing, where the model is the composition of the inferred models,
and the IUT is the composition of the components. 
If a fault is discovered during this phase, it is used as a counter-example of a
conjecture query, and a new inference phase is started.

There are still open issues with this method. 
It terminates when a model-based testing phase has found no fault after achieving a
given coverage criteria of the current combined model: thus, there is no assessment
of the approximation reached by the inferred models, which is dependent of the
choice of the criteria, and there is no guarantee of termination. 
Moreover, performing model-based testing on such global models may lead to state
explosion, and may be beyond the current state of the art.

\subsubsection{Coverage-biased random testing}\label{crt}

In presence of very large models, drawing at random checking sequences is one of the
practical alternatives to their systematic and exhaustive construction, as presented
in Section \ref{test}.

\begin{figure}[h]
\centerline{\includegraphics[height=4.2cm]{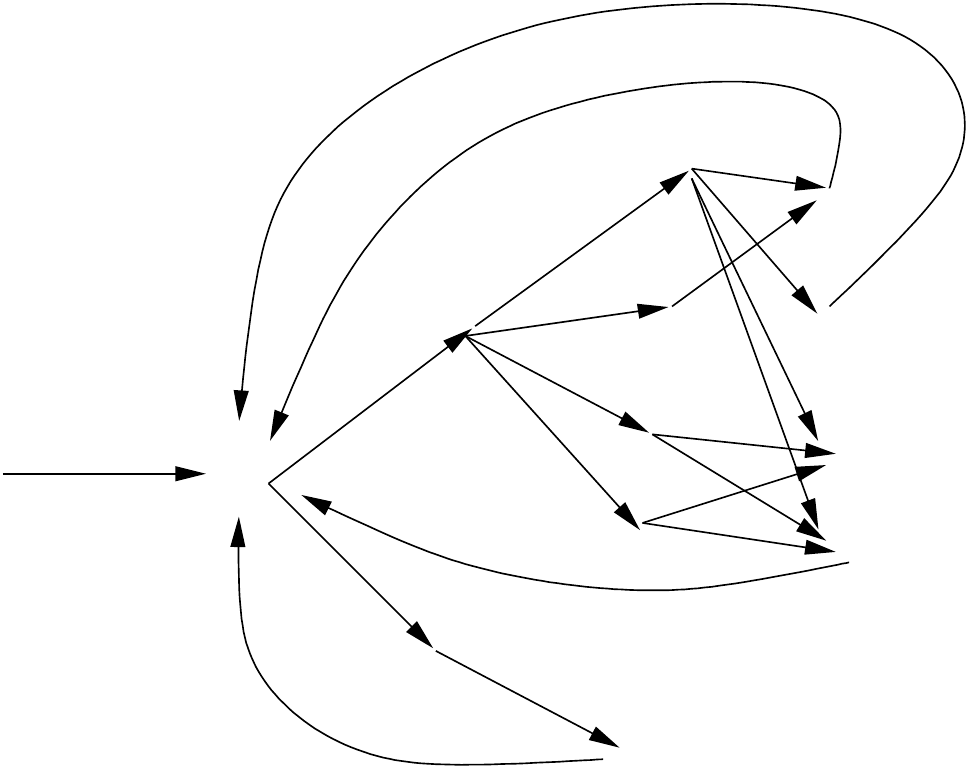}}
\caption{Irregular topology for which classical random walks is not
uniform.\label{mcg}}
\end{figure}

Testing methods based on random walks have already been mentioned in Section \ref{pst}.
However, as noted in \cite{SG03}, classical random
walk methods have some drawbacks.  In case of irregular topology of
the underlying transition graph, uniform choice of the next state is
far from being optimal from a coverage point of view (see Figure~\ref{mcg}). 
Moreover, for
the same reason, it is generally not possible to get any estimation of
the test
coverage obtained after one or several random walks: it would require
some complex global analysis of the topology of the model.

One way to
overcome these problems has been proposed for
program testing in \cite{GDG01,DGG04,D12}, and is applicable to model-based testing.  It
relies upon techniques for
counting and drawing uniformly at random combinatorial structures seen in Section
\ref{counting}.

The idea of \cite{GDG01,DGG04,D12} is to give up, in the random walk,
the uniform choice of the next state and to bias this choice according
to the number of elements (traces, or states, or transitions)
reachable via each successor.  The estimation of the number of traces ensures
 a  uniform probability on traces.  Similarly by considering
 states or transitions, it is possible to maximize
the minimum probability to reach such an element.
Counting the traces starting from a given state, or those traces traversing specific
elements can be efficiently performed with the methods of Section \ref{counting}.

Let $ D $ be some description of a system under test.
$ D $ may be a model or a program, depending on the kind of test
we are interested in (black box or structural).
We assume that $ D $ is based on a graph (or a tree, or more generally, on some
kind of combinatorial structure).
On the basis of this graph, it is possible to define coverage criteria:
all-vertices, all-edges, all-paths-of a certain-kind, etc.
More precisely, a coverage criterion $ C $ characterizes for a given
description $ D $ a set of elements $ E_C(D) $ of the underlying graph
(noted $ E $ in the sequel when $ C $ and $ D $ are obvious).
In the case of deterministic testing, the criterion is satisfied by a test suite if
every
element of the $ E_C(D) $ set is reached by at least one test. 

In the case of random testing, the notion of coverage must be revisited. 
There is some distribution $\Omega$ that is used to draw tests (either input
sequences or traces).
Given $ \Omega$,   the satisfaction of a coverage
criteria $ C $ by a testing method for a description $ D $  is characterized by
the minimal probability $ q_{C,N}(D) $ of covering any element of $ E_C(D)
$ when drawing $ N $ tests.
In \cite{Th89}, Thevenod-Fosse and Waeselink called $ q_{C,N}(D) $ the test quality
of the method
with respect to $ C $.

Let us first consider a method based on drawing at random paths in a finite subset
of them (for instance
 $ {\cal{P}}_{\leq n} $, the set of paths of length less or equal to $n$), and on
the coverage criteria $C$ defined by this subset.
As soon as the test experiments are independent, 
this test quality $ q_{C,N}(D) $ can be easily stated provided that $ q_{C,1}(D) $
is known.
Indeed, one gets $ q_{C, N} (D) = 1 -(1 - q_{C,1}(D))^N $.

The assessment of test quality is more complicated in general.
Let us now consider more practicable coverage criteria, such as
``all-vertices'' or ``all-edges'', and some given random testing
method.
Uniform generation of paths does not ensure optimal quality when the elements of
$E_C(D)$ are not
paths, but are constitutive elements of the graph as, for example, vertices, edges,
or cycles.
The elements to be covered generally have different probabilities to be
reached by a test:
some of them are covered by all the tests,
some of them may have a very weak probability, due to the structure of the
behavioral graph or to some specificity of the testing method.

Let $ E_C(D) = \{ e_1, e_2, ..., e_m \} $ and for any $  i \in \{1,...,m\},  p_i
$ the probability for the element $ e_i $ to be exercised during the
execution of a test generated by the considered testing
method. Let $p_{min} = min \{p_i | i \in   \{1,...,m\}\}$. 
Then
\begin{equation}
q_{C,N} (D) \ge 1 - (1 - p_{min} )^N 
\label{equationPmin}
\end{equation}

Consequently, the number $ N $ of tests required to reach a given quality 
$ q_C (D) $ is

$$ N \geq  \frac{log(1 - q_C(D))}{log(1 - p_{min})} $$

By definition of the test quality, $ p_{min}$ is just $q_{C,1}(D)$. 
Thus, from the formula above one immediately deduces that for any given $ D $,
for any given $ N $, maximizing the quality of a random testing method with respect
to a
coverage criteria $ C $ reduces to maximizing $ q_{C,1}(D) $, i. e. $ p_{min} $.
In the case of random testing based on a distribution $\Omega$,  $ p_{min} $
characterizes, for
a given coverage criteria $C$, the approximation of the coverage induced by $\Omega$.

However, maximizing $p_{min}$ should not lead to give up the randomness of the method.
This may be the case when there exists a path traversing all the elements of $E_C(D)$: 
one can maximize $p_{min}$ by giving a probability 1 to this path, going back to a
deterministic
testing method. Thus, another requirement must be combined to the maximization of
$p_{min}$: all the paths traversing an element of $E_C(D)$ must have a non null
probability and the minimal
probability of such a path must be as high as possible.
Unfortunately, these two requirements are antagonistic in many cases.

In \cite{GDG01,DGG04,D12}, the authors propose a practical solution in two steps:
\begin{enumerate}
\item pick at random an element $e$ of $E_C(D)$, according 
to a suitable probability distribution (which is discussed below);
\item generate uniformly at random a path of length $\leq n$ 
that goes through $e$. (This ensures a balanced coverage of the 
set of paths which cover $e$.) 
\end{enumerate}

Let  $\pi_i$ the probability of choosing element $e_i$ in step~1 of the process above.

Given $\alpha_i$ the number of paths of ${\cal P}_{\leq n}$, which cover 
element $e_i$,
given $\alpha_{i,j}$ the number of paths, which cover both elements $e_i$ and 
$e_j$; (note that $\alpha_{i,i}$ = $\alpha_{i}$
and $\alpha_{i,j}$ = $\alpha_{j,i}$),
the probability of reaching
$e_i$ by drawing a random path which goes through another element
$e_j$ is $\frac {\alpha_{i,j}} {\alpha_j}$.  
Thus the probability $p_i$
for the element $e_i$ (for any $i$ in $(1..m)$) to be reached by a
path is
$$ p_i = \pi_i + \sum_{j \in (1..m)-\{i\}} \pi_j {\frac
{\alpha_{i,j}}{\alpha_j}}\,, $$

The above equation simplifies to
\begin{equation} \label{eqpi}
  p_i = \sum_{j=1}^{m} \pi_j {\frac {\alpha_{i,j}} {\alpha_j}}
\end{equation}
since $\alpha_{i,i}=\alpha_i$. Note that coefficients $\alpha_j$ and
$\alpha_{i,j}$ are easily computed by ways given in
Section \ref{counting}.

The determination of the probabilities $\{\pi_1, \pi_2, \ldots,
\pi_m\}$ with $\sum \pi_i = 1$, which 
maximize $p_{min}=\min\{p_i, i \in  \{1,...,m\} \}$ can be stated as a
linear programming problem:
\begin{center}
Maximize $p_{min}$ under the constraints: 
$
\left\{
\begin{array}{l}
\forall i\leq m,~~~ p_{min} \leq p_i \;; \\
\pi_1 + \pi_2 + \cdots + \pi_m = 1\;;
\end{array}
\right.
$
\end{center}
where the $p_i$'s are computed as in Equation~(\ref{eqpi}). Standard
methods lead to a solution in time polynomial according to $m$.

Starting with the principle of a two-step drawing strategy, first an element in
$E_C(D)$, second a path
among those  traversing this element,
this approach ensures a maximal minimum probability of reaching the elements to be
covered and, once this
element is chosen, a uniform coverage of the paths traversing this element.
For a given number of tests, it makes it possible to assess the approximation of the
coverage,
and conversely, for a required approximation, it gives a lower bound of the number
of tests to
reach this approximation.

The idea of biasing randomized test methods in function of a coverage criterion was
first
studied in the nineties  in \cite{ThWa91}, but the difficulties
of automating the proposed methods prevented their exploitation.
More recently, this idea has been explored also in the Pathcrawler and Dart tools
\cite{Pathcrawler,Dart}, with a limitation to coverage criteria based on paths.

\subsection{Approximate probabilistic model checking \label{APMC}}

The main approaches to reduce the prohibitive space cost of probabilistic model checking try to generalize predicate abstraction coupled with counterexample-guided abstraction refinement (CEGAR) to a probabilistic setting. An approach to develop probabilistic CEGAR \cite{hwz} is based on the notion and the interpretation of counterexamples in the probabilistic framework. A quantitative analog of the well-known CEGAR loop is presented in \cite{kknp}. The underlying theory is based on representing abstractions of Markov Decision Processes as two-player stochastic games. The main drawback of these approaches is that the abstraction step that is repeated during the abstraction refinement process does not ensure a significant gain, i.e. exponential, in terms of space.
We present now an other approximation method for model checking probabilistic transition 
systems. This approach uses only a succinct representation of the model to check, i.e. a program describing the probabilistic transition system in some input language of the model checker. Given some probabilistic transition system and some linear
temporal formula $\psi$, the objective is to approximate $Prob[\psi]$ by using probabilistic algoritms whose complexity is logspace.
There are serious complexity reasons to think that 
one cannot efficiently approximate this probability for  a general LTL
formula. However, if the problem is restricted to an LTL fragment sufficient to express interesting properties such than reachability and safety, one can obtain efficient approximation algorithms.

\subsubsection{Probability problems and approximation}

The class $\# P$ captures the problems of counting
the numbers of solutions to $NP$ problems. The counting versions of
all known $NP$-complete problems are $\# P$-complete. The well
adapted notion of reduction is parsimonious reduction: it is a
polynomial time reduction from the first problem to the second one,
recovering via some oracle, the number of solutions for the first
problem from the number of solutions for the second one.
Randomized versions of
approximation algorithms exist for problems such as counting the
number of valuations satisfying a propositional disjunctive normal
form formula ($\# DNF$) \cite{kl2} or network reliability problem
\cite{k}. But we remark that it does not imply the existence of FPRAS
for any $NP$-complete problem.

A probability problem is defined by giving as input a 
representation of a probabilistic system and a property, as output
the probability measure $\mu$ of the measurable set of execution
paths satisfying this property.
 One can adapt the notion of fully polynomial randomized
approximation scheme, with multiplicative or additive error,
to probability problems. In the following theorem, $RP$
is the class of decision problems that admit one-sided error 
polynomial time randomized algorithms.

\begin{theorem}
There is no fully polynomial randomized approximation scheme (FPRAS) for the
  problem of computing $Prob[\psi]$ for $LTL$ formula $\psi$, unless
  $RP=NP$.
\end{theorem}

In the following, we give some idea of the proof.
We consider the fragment $L(\mathbf{F})$ of $LTL$ in which
 $\mathbf{F}$ is the only temporal operator.
The following result is due to Clarke and Sistla \cite{sc}: the
problem of deciding the existence of some path satisfying a
$L(\mathbf{F})$ formula in a transition system is $NP$-complete.
Their proof uses a polynomial time reduction of $SAT$ to the problem
of deciding satisfaction of $L(\mathbf{F})$ formulas. From this
reduction, we can obtain a one to one, and therefore parsimonious,
reduction between the counting version of $SAT$, denoted by $\# SAT$,
 and counting finite paths, of given length, whose extensions satisfy the associated
$L(\mathbf{F})$ formula.

A consequence of this result is the $\# P$-hardness of
computing satisfaction probabilities for general $LTL$ formulas.
We remark that if there was a FPRAS for approximating $Prob[\psi]$ for
$LTL$ formula $\phi$, we could efficiently approximate $\# SAT$.
A polynomial randomized approximation scheme for $\# SAT$ could be
used to distinguish, for input $y$, between the case $\# (y) =0$
and the case $\# (y) >0$, thereby implying a randomized polynomial
time algorithm for the decision version $SAT$.
 
As a consequence of a result of  \cite{jvv} 
and a remark of 
 \cite{bk:Sinclair}, the existence of an FPRAS for $\# SAT$ would imply $RP=NP$.
On the other hand, $\# SAT$ can be approximated with an additive
error by a fully polynomial time randomized algorithm.
In the next section, we determine some restriction on the class of linear
temporal formulas $\psi$, on the value $p = Prob[\psi]$ and only consider
approximation with additive error in order to
obtain efficient randomized approximation schemes for  such 
probabilities.

\subsubsection{A positive fragment of LTL}

For many natural properties, satisfaction on a path of
length $k$ implies satisfaction by any extension of this path.  Such
properties are called monotone.  Another important
class of properties, namely safety properties, can be expressed as
negation of monotone properties. One can reduce the computation
of satisfaction probability of a safety property to the same problem
for its negation, that is a monotone property.
Let consider a subset of LTL formulas
which allows to express only monotone properties and for which one can
approximate satisfaction probabilities.

\begin{definition}
  The {\em essentially positive fragment} (EPF) of LTL is the set of
  formulas constructed from atomic formulas ($p$) and their negations
  ($\neg p$), closed under $\vee$, $\wedge$ and the temporal operators
  $\mathbf{X, U}$.
\end{definition}

For example, formula $\mathbf{F}p$, that expresses a reachability
property, is an $EPF$ formula. 
Formula $\mathbf{G}p$, that expresses a safety
property, is equivalent to $\neg \mathbf{F} \neg p$, which is the negation
of an $EPF$ formula. 
Formula $\mathbf{G} (p \rightarrow \mathbf{F}q)$,
that expresses a liveness property, is not an $EPF$ formula, nor
equivalent to the negation of an $EPF$ formula.
In order to approximate the satisfaction probability $Prob [\psi]$ of
an $EPF$ formula, let first consider $Prob_k [\psi]$, the probability
measure associated to the probabilistic space of execution paths of finite
length $k$.
The monotonicity of the property defined by an $EPF$ formula gives the following
result.

\begin{proposition}
  Let $\psi$ be an LTL formula of the essentially positive fragment
  and ${\cal M}$ be a probabilistic transition system.  Then the
  sequence $(Prob_k [\psi])_{k \in \mathbb{N}}$ converges to $Prob
  [\psi]$.
\end{proposition}

A first idea is to approximate $Prob_k[\psi]$ and to use a fixed point
algorithm to obtain an approximation of $Prob [\psi]$. This
approximation problem is believed to be intractable for deterministic
algorithms. In the next section, we give a randomized approximation
algorithm whose running time is polynomial in the size of a succinct
representation of the system and of the formula. Then we deduce a randomized
approximation
algorithm to compute $Prob[\psi]$, whose space complexity is logspace.

\subsubsection{Randomized approximation schemes}\label{sras}

\paragraph{Randomized approximation scheme with additive error.}
We show that one can approximate the satisfaction probability of an
$EPF$ formula with a simple randomized algorithm.  In practice
 randomized approximation with additive error is
sufficient and gives simple algorithms, we first explain how to design it. 
Moreover, this randomized approximation is fully polynomial for
bounded properties. 
Then we will use the estimator theorem \cite{kl2} and an optimal
approximation algorithm \cite{dagum} in order to obtain randomized
approximation schemes with multiplicative error parameter, according to
definition \ref{fpras}. 
In this case the
randomized approximation is not fully polynomial even for bounded properties.

One generates random paths in the probabilistic space underlying the
Kripke structure of depth $k$ and computes a random variable $A$ which
additively approximates $Prob_k [\psi]$. This approximation will be
correct with confidence $(1-\delta)$ after a polynomial number of
samples.
 The main advantage of the method is that one can proceed with just a
succinct representation of the transition system, that is a succinct
description in the input language of a probabilistic model checker as PRISM. 

\begin{definition}
  A succinct representation, or diagram, of a $PTS$ ${\cal M}=(S, s_0, M ,L)$ is a representation of the $PTS$, that allows to generate
  for any state $s$, a successor of $s$  with respect to the
  probability distribution induced by $M$.
\end{definition}

The size of such a succinct representation is substantially smaller than
the size of the corresponding $PTS$.  Typically, 
the size of the diagram is polylogarithmic in the size of the $PTS$, thus
eliminating the space complexity problem due to the state space explosion
phenomenon.
The following function \textbf{Random Path} uses such a succinct representation to
generate a 
random path of length $k$, according to the probability matrix $P$, and to check the
formula $\psi$:

\begin{center}
\begin{fmpage}{10cm}
\textbf{Random Path}\\
\textbf{Input: } $diagram_{\cal M},k,\psi$\\
\textbf{Output: } samples a path $\pi$ of length $k$ and check formula $\psi$ on $\pi$
\begin{enumerate}
\item Generate a random path $\pi$ of length $k$ (with the diagram)
\item If $\psi$ is true on $\pi$ then return $1$ else $0$
\end{enumerate}
\end{fmpage}
\end{center}

Consider now the random sampling algorithm $\mathcal{GAA}$ designed
for the approximate computation of $Prob_{k} [\psi]$:

\begin{center}
\begin{fmpage}{10cm}
\textbf{Generic approximation algorithm ${\cal GAA}$}\\
\textbf{Input: } $diagram_{\cal M},k,\psi,\epsilon,\delta$\\
\textbf{Output: } approximation of $Prob_k [\psi]$\\
$N:=\ln(\frac{2}{\delta})/ 2 \epsilon ^2$\\
$A:=0$\\
For $i=1$ to $N$ do
 $A:=A+ \textbf{Random Path}(diagram_{\cal M},k,\psi)$
Return $A/N$
\end{fmpage}
\end{center}

\begin{theorem}\label{ras}
  The generic approximation algorithm ${\cal GAA}$ is a fully
  polynomial randomized approximation scheme (with additive error
  parameter) for computing $p=Prob_k [\psi] $ whenever $\psi$ is in
  the $EPF$ fragment of $LTL$ and $p \in ]0,1[$.
\end{theorem}

One can obtain a randomized approximation of $Prob[\psi]$ by iterating
the approximation algorithm described above. Detection of time
convergence for this algorithm is hard in general, but can be characterized for
the important case of ergodic Markov chains. The logarithmic
 space complexity is an important 
feature for
applications.

\begin{corollary}
  The fixed point algorithm defined by iterating the approximation
  algorithm $\mathcal{GAA}$ is a randomized approximation scheme, whose
  space complexity is logspace, for the probability problem $p=Prob
  [\psi] $  whenever $\psi$ is in the $EPF$ fragment of $LTL$ and $p \in ]0,1[$.
\end{corollary}

For ergodic Markov chains, the convergence rate of  $Prob_k [\psi]$ to
$Prob[\psi]$ is in $O(k^{m-1} |\lambda|^k)$ where $\lambda$ is the
second eigenvalue of $M$ and $m$ its multiplicity.
The randomized approximation algorithm described above is implemented
in a distributed probabilistic model checker named APMC \cite{DBLP:conf/qest/HeraultLP06}. Recently this
tool has been extended to the verification of continuous time Markov chains.

\paragraph{Randomized approximation scheme with multiplicative error.} We use a
generalization of the zero-one estimator theorem \cite{kl2}
to estimate the expectation $\mu$ of a random variable $X$ distributed
in the interval $[0,1]$. 
The generalized zero-one estimator theorem
\cite{dagum} proves that if $X_1, X_2,\dots, X_N$ are random variables
independent and identically distributed according to $X$, $S =
\sum_{i=1}^N X_i$, $\epsilon <1$, and $N =
4(e-2). \ln(\frac{2}{\delta}) . \rho / (\epsilon . \mu)^2$, then $S
/N$ is an $(\epsilon ,\delta)$-approximation of $\mu$, i.e.:
$$ Prob \big(\mu (1- \epsilon) \leq S/N \leq \mu (1+ \epsilon) \big)
\geq 1 - \delta $$ where $\rho = \textit{max} (\sigma^2 , \epsilon
\mu)$ is a parameter used to optimize the number $N$ of experiments
and $\sigma^2$ denotes the variance of $X$.
In \cite{dagum}, an optimal approximation algorithm, running in three
steps, is described:

\begin{itemize}
\item using a stopping rule, the first step outputs an $(\epsilon
,\delta)$-approximation $\hat{\mu}$ of $\mu$ after an expected number of
experiments proportional to $\Gamma / \mu$ where $\Gamma =
4(e-2). \ln(\frac{2}{\delta})/ \epsilon^2$;
\item the second step uses the value of $\hat{\mu}$ to set the number
of experiments in order to produce an estimate $\hat{\rho}$ that is
within a constant factor of $\rho$ with probability at least $(1 -
\delta)$;
\item the third step uses the values of $\hat{\mu}$ and $\hat{\rho}$
to set the number of experiments and runs these experiments
to produce an $(\epsilon ,\delta)$-approximation of $\mu$.
\end{itemize}

One obtains a randomized approximation scheme with multiplicative error
by applying the optimal approximation algorithm $\mathcal{OAA}$ with
input parameters $\epsilon ,\delta$ and the sample given by the
function \textbf{Random Path} on a succinct representation of
$\mathcal{M}$, the parameter $k$ and the formula $\psi$.

\begin{theorem}\label{tfpras}
  The optimal approximation algorithm ${\cal OAA}$ is a 
  randomized approximation scheme (with multiplicative
  error) to compute $p=Prob_k [\psi] $ whenever $\psi$ is
  in the $EPF$ fragment of $LTL$ and $p \in ]0,1[$.
\end{theorem}
 
We remark that the optimal approximation algorithm is not an $FPRAS$
as the expected number of experiments $\Gamma / \mu$ can be
exponential for small values of $\mu$. 

\begin{corollary}
  The fixed point algorithm defined by iterating the optimal
  approximation algorithm ${\cal OAA}$ is a randomized approximation
  scheme for the probability problem $p=Prob [\psi] $ whenever $\psi$
  is in the $EPF$ fragment of $LTL$ and $p \in ]0,1[$.
\end{corollary}

\section{Conclusion}
Model checking and testing are two areas with a similar goal: to verify that a
system satisfies a property. They start with  different hypothesis on the systems
and develop many techniques with different notions of  approximation, when an exact
verification may be  computationally too hard. 

We presented some of the well known notions of approximation with their logic and
statistics backgrounds, which yield several techniques for model checking and
testing. These methods guarantee the quality and the efficiency of the approximations.

\begin{enumerate}
\item In bounded model checking, the approximation is on the length of the computation paths to witness possible errors, and
 the method is polynomial in the
size of the model.
\item In approximate model checking, we developped two approaches. In the first one, the approximation is  on the density of errors
and the Monte Carlo methods are polynomial in the size of the model. In the second one, the approximation is  on
the distance of the inputs and the complexity of the property testers is independent of the size of the model and only dependent on 
$\eps$.

\item In approximate black box checking, learning techniques construct a model which can be compared with a property in 
exponential time. 
The previous approximate model checking technique
  guarantees that the  model is
$\eps$-close to the IUT after $N$ samples, where $N$ only depends on $\eps$.

\item In approximate model-based testing, a coverage criterium is satisfied with
high probability and the method is polynomial in
the size of the representation.

\item In approximate probabilistic model checking,  the estimated probabilities of
satisfying formulas are close to the real ones. The method is polynomial in the size
of the given succinct representation.
\end{enumerate}

Some of these approximations can be combined for future research. For example, approximations
used in black box checking and model-based testing can be merged, as learning
methods influence the new possible tests. As another example, probabilistic model
checking and approximate model checking can also be merged, as we may decide if a
probabilistic system is close to satisfy a property.

\bibliographystyle{alpha}
\bibliography{reference1,mcgbib07,lassaignebib07,MCG120308}
\end{document}